 \documentclass[prodmode,acmcsur]{acmsmall}

\usepackage[ruled]{algorithm2e}

\SetAlFnt{\small}
\SetAlCapFnt{\small}
\SetAlCapNameFnt{\small}
\SetAlCapHSkip{0pt}
\IncMargin{-\parindent}

\usepackage{amsmath, amssymb, epsfig, psfrag}
\usepackage{multirow}
\usepackage{sidecap}
\usepackage{subfigure}
\usepackage{array}
\usepackage{colortbl}
\usepackage{url}
\usepackage{cite}

\acmArticle{1}
\acmYear{2013}
\acmMonth{2}

\def\beq{\begin{equation}}
\def\eeq{\end{equation}}
\def\beqa{\begin{eqnarray}}
\def\eeqa{\end{eqnarray}}
\def\beqan{\begin{eqnarray*}}
\def\eeqan{\end{eqnarray*}}

\def\qed{\mbox{} \hfill $\Box$}
\begin{document}

\markboth{S. Sen et al.}{A Survey of Smart Data Pricing: Past Proposals, Current Plans, and Future Trends}

\title{A Survey of Smart Data Pricing:\\ Past Proposals, Current Plans, and Future Trends}
\author{SOUMYA SEN		
\affil{Princeton University}
CARLEE JOE-WONG
\affil{Princeton University}
SANGTAE HA
\affil{Princeton University}
MUNG CHIANG
\affil{Princeton University}  
}

%
%

\begin{abstract}
Traditionally, network operators have used simple flat-rate broadband data plans for both wired and wireless network access. But today, with the popularity of mobile devices and exponential growth of apps, videos, and clouds, service providers are gradually moving towards more sophisticated pricing schemes. This decade will therefore likely witness a major change in the ways in which network resources are managed, and the role of economics in allocating these resources.  This survey reviews some of the well-known past broadband pricing proposals (both static and dynamic), including their current realizations in various consumer data plans around the world, and discusses several research problems and open questions. By exploring the benefits and challenges of pricing data, this paper attempts to facilitate both the industrial and the academic communities' efforts in understanding the existing literature, recognizing new trends, and shaping an appropriate and timely research agenda. 
\end{abstract}



 
\acmformat{Sen, S., Joe-Wong, C., Ha, S., Chiang, M. 2013. A Survey of Smart Data Pricing: Past Proposals, Current Plans, and Future Trends}

\begin{bottomstuff}
This work is in part supported by the National Science Foundation, under
grant NSF CNS-1117126, and C. J.-W.'s NDSEG fellowship.

Authors' addresses: S. Sen {and} S. Ha {and} M. Chiang, Electrical Engineering Department,
Princeton University;  C. Joe-Wong, Program in Applied \& Computational Mathematics, Princeton University.
\end{bottomstuff}

\maketitle

\section{Introduction}\label{intro}


In November 1996, AOL, the largest U.S. Internet Service Provider\footnote{An \emph{Internet Service Provider} is defined as a company offering its customers subscriptions for wired and/or wireless access to the Internet, e.g., AT\&T, Comcast, and Verizon in the United States.} (ISP) of the time, switched from hourly pricing to a flat-rate monthly data plan of  \$19.95 \cite{Lewis}.  One week later, Pacific Telesis Enterprises CEO M. Fitzpatrick made an apocalyptic prediction \cite{Marshall}:
\begin{quote}
\emph{My indicators point to a genuine data tsunami, a tidal wave, in the next 18 months. And, ladies and gentlemen, while we can surf the net, nobody can surf a tidal wave.}
\end{quote}

Fortunately, that devastating tsunami did not come, but the tides of growth in the demand for data did help the Internet grow.  

However, today, the fear of a data tsunami is slowly coming back \cite{att-tsunami}.
The Cisco Visual Networking Index \cite{CiscoVNI} predicted in May 2012 that global IP (Internet Protocol) traffic will reach 1.3 zettabytes/year (1 zettabyte = $10^{21}$ bytes) and mobile data will reach 798 exabytes/year (1 exabyte = $10^{18}$ bytes) by the end of 2016.  
With the growing popularity of iPhones, iPads, bandwidth-hungry applications, and cloud-based services, ISPs have been increasingly turning to pricing as the ultimate congestion management tool to mitigate this growth --a feat often achieved by imposing harsh overage penalties and ``throttling'' the very customers who drive this demand.

The basic idea of congestion-based pricing has existed for several decades in many different utility markets, e.g., transportation and electricity distribution networks. As for the Internet, even as early as 1997, when it was just evolving into a commercial service and its revenue models were hotly debated, MacKie-Mason et al. \citeyear{MacKie-Mason} wrote:
\begin{quote}\emph{We argue that a feedback signal in the form of a variable price for network service is a workable tool to aid network operators in controlling Internet traffic. We suggest that these prices should vary dynamically based on the current utilization of network resources.}
\end{quote}
They were proposing a form of dynamic congestion pricing for the Internet.  As a mathematically oriented research topic, congestion pricing has been extensively studied in transportation networks, energy markets, telephone networks, ATM and IP networks, etc. Thus, in some sense, there are not many new statements to be made about the generic theory of congestion pricing. And yet, wireless and wireline data networks have so far used the most rudimentary forms of congestion pricing, e.g., usage-based charges.  This can partly be explained by Clark's \citeyear{DDClark} observation:
\begin{quote}
\emph{Whatever the providers may prefer to do, competition may force some forms of fixed-fee pricing in the marketplace.}
\end{quote}
While providers may have preferred flat rates in the 1990s, the situation is quite different now: the problem has worsened and operators are more aggressive in pursuing new pricing schemes. The acceptance of this fact is perhaps nowhere more clearly evident than in the ``New Rules for an Open Internet"  \cite{FCC-JG} announced on December 21, 2010, by Federal Communications Commission (FCC) chairman J. Genachowski:
\begin{quote}
\emph{The rules also recognize that broadband providers need meaningful flexibility to manage their networks to deal with congestion.... And we recognize the importance and value of business-model experimentation.}
\end{quote}
Therefore, the interesting question is: given the rapid rise in capacity demand in the age of apps and clouds, how will pricing policies change over this decade? 

The shift in pricing trends is more easily noticeable in growing economies, e.g., India and Africa, where dynamic congestion pricing for voice calls is already practiced.  The next logical step in this pricing evolution is dynamic pricing for data.  But pricing data has several unique challenges (and opportunities), arising from both technological and social considerations.  A clear understanding of these issues requires consideration of both the past and the present developments. To this end, this paper reviews some of the best-known pricing proposals of the last two decades and reports on some very interesting pricing schemes that are in use today, along with the identification of research challenges and emerging trends in access pricing.



\subsection{Contributions}\label{contrib}

In 1996, Breker \citeyear{Breker} conducted a survey of computer network pricing schemes that covered a few of the known protocols of the time.  Similar reviews of pricing concepts for broadband IP and integrated services networks were undertaken about a decade ago by Falkner et al. \citeyear{Falkner} and Chang and Petr \citeyear{chang2001survey}. While these surveys mentioned many concerns that exist even today--e.g., the need to manage network congestion--since then, researchers have proposed several new schemes for pricing data, and some variations of these plans have been adopted by wireless ISPs for their consumers. Ezziane \citeyear{Ezziane} surveyed the charging and billing mechanisms used by 3G wireless service providers. In particular, Ezziane reported on new approaches needed for mediation, billing, and charging for these services.  More recently, Gizelis and Vergados \citeyear{Gizelis} provided a detailed overview of the literature on pricing schemes for wireless networks, while Sen et al. \citeyear{sen2012incentivizing} studied pricing approaches that incentivize time-shifting of demand. An annotated bibliography of several papers related to various aspects of Internet economics, including pricing, was compiled by Klopfenstein \citeyear{Klopfenstein}.  Our work not only complements these earlier surveys and reports on broadband pricing, 
but also attempts to understand the kinds of pricing plans currently offered by wired and wireless ISPs around the world, while noting some of the weaknesses in their approaches.  Additionally, we provide a detailed discussion on the various technological, socio-economic, and regulatory challenges to pricing innovation, and identify several directions for future research. Our focus in this article is on the pricing for end-user/consumer rather than the economics of peering or transit-fee agreements between service providers. The key features of this work are summarized below:
\begin{itemize}
\item Identification of the threats to the Internet ecosystem from the perspective of ISPs, consumers, and content providers.
\item Detailed analysis of the various challenges to access pricing innovation and open problems for future research.
\item A broad overview of several past and recent proposals for pricing data, along with a classification of static and dynamic pricing plans.
\item Several illustrative examples of real pricing plans implemented by ISPs in different parts of the world.
\item An overview of the relationships between the ideas of broadband data pricing and earlier congestion pricing ideas in road and electricity networks.
\item Discussion on the other directions in access pricing innovation, including satellite and heterogeneous networks.
\end{itemize}
 
Networking researchers and operators, caught in their conscientious efforts to find the best solutions to this growing problem, often tend to overlook many of the innovative practices already deployed by network operators in the real-world. However, appraising the past several years of theoretical foundations on pricing and identifying their varied realizations in the present world will likely be key to predicting future trends and shaping an appropriate research agenda. The information presented in this work draws from a wide range of sources, including research proposals, existing data plans, news articles, consumer forums, and reports from experimental field trials. 

This paper is organized as follows: Section \ref{sec:threats} provides an overview of threats to the Internet ecosystem from network congestion.  An overview of the role of pricing in achieving various network functionality goals, the fundamentals of network economics, and the potential benefits of access pricing schemes in alleviating network congestion is provided in Section \ref{sec:pricing_data_networks}.  Section \ref{sec:challenges} identifies the key challenges to overcome in order to innovate access pricing and outlines directions for future research.  A review of the proposals and realizations of several static and dynamic pricing plans is provided in Sections \ref{sec:static} and \ref{sec:dynamic}, respectively.  The emerging trends in network pricing evolution and their roles in satellite and heterogeneous networks are addressed in Section \ref{sec:directions}. Section \ref{conclusions} concludes with a summary of this work.

\section{Threats to the Internet Ecosystem}\label{sec:threats}

The Internet ecosystem today suffers from threats to both its \emph{sustainability} and \emph{economic viability}. The sustainability concern is due to the near doubling of demand for mobile data every year  \cite{CiscoVNI}, which brings into question the feasibility of finding any purely technical solutions to this problem. Additionally, changes in the policy to make more spectrum available, although useful, is unlikely to provide any long-term solution.  For example, the US Government's initiative in releasing an additional 50 MHz of spectrum into the market for six carriers to compete for is widely viewed as being quite insufficient \cite{tollfree}. 

The latter concern regarding the economic viability of Internet access is due to the large overage charges and throttling that ISPs (e.g., AT\&T and Verizon) are now imposing on their customers to penalize demand. To understand the implications of these factors, we need to first look at the impact that this growth is having on the various stakeholders of the Internet ecosystem before delving deeper into the promises that \emph{Smart Data Pricing} (SDP) holds \cite{SDP}. In the following discussion, we explore the current challenges and threats from the perspective of carriers, consumers, and content providers.

\subsection{ISPs' Traffic Growth}    

The Cisco Visual Networking Index \citeyear{CiscoVNI} predicted in 2012 that mobile data traffic\footnote{``Mobile data traffic'' includes handset-based data traffic, such as text messaging, multimedia messaging, and handset video services, as well as traffic generated by notebook cards and mobile broadband gateways.} will grow at a compound annual growth rate (CAGR) of 78\% between 2011 and 2016, reaching a volume of 10.8 exabytes per month in 2016. By that time, Cisco predicted that the average smartphone user will consume 2.6 GB per month, as opposed to 150 MB/month in 2011 \citeyear{CiscoVNImobile}. This growth in per-device data consumption is fueled by the demand for bandwidth-intensive apps: for example, Allot Communications reports that usage of Skype's video calling service grew by 87\% in 2010, while usage of Facebook's mobile app grew by 267\% \cite{allot}. Devices themselves are also contributing to the increase in traffic volume; Apple's iPad 3 quadrupled the screen resolution from the iPad 2, allowing videos of higher quality to be streamed to the device \cite{apple}.

Cisco predicted that by 2016, Wi-Fi and mobile data will comprise 61\% of all Internet traffic, with wired traffic comprising the remaining 39\% \cite{CiscoVNI}. Although the faster growth rate in mobile data traffic is a major concern, growing congestion on wired networks both at the edge and in the networks' middle mile also poses a significant problem. 
By 2016, ISPs are expected to carry 18.1 petabytes per month in managed IP traffic.\footnote{Cisco's definition of ``managed IP'' includes traffic from both corporate IP wide area networks and IP transport of television and video-on-demand.} That this growth is causing concern among ISPs can be seen from Comcast's recent initiative to cap their wired network users to 300 GB per month \cite{Comcast-cap}. Indeed, even in 2008 Comcast made headlines with their (since reversed) decision to throttle Netflix as a way to curb network congestion \cite{Comcast-Level3}. Video streaming due to services like Netflix is in fact a major contributor to wired network traffic--by 2016, Cisco predicts that fixed IPs will generate 40.5 petabytes of Internet video per month \citeyear{CiscoVNI}.
  
Congestion in wired networks also affects rural local exchange carriers (RLECs), due to the persistence of the middle-mile problem for RLECs. While the cost of middle mile bandwidth has declined over the years due to an increase in the DSL demand needed to fill the middle mile, the bandwidth requirements of home users have increased sharply \cite{vglass-2}. The FCC has set a broadband target rate of 4 Mbps downstream speed for home users, but the average speed provided to rural customers is much lower. The cost of middle mile upgrades to meet the target speeds will be substantial and could be a barrier to providing greater speeds for subscribers and digital expansion in the rural areas \cite{vglass-2}. Therefore, research on access pricing to bring down middle mile investment costs by reducing the peak capacity and RLECs' over-provisioning needs will be a crucial step in bridging the digital divide.

\subsection{Consumers' Cost Increase}

With the growth in the volume of Internet traffic, ISPs have begun to pass some of their network costs on to consumers. Consumers, on the other hand, have become increasingly concerned at the high costs of Internet subscriptions. For instance, when AT\&T and Verizon announced in July 2012 that they were offering shared data plans for all new consumers and discontinuing their old plans, many consumers ended up with higher monthly bills \cite{nyt-shared}. Users, in fact, are aware of these increasing costs; while smartphone adoption is still growing in the United States, so is the popularity of usage-tracking and data compression apps that help mobile users stay within their monthly data caps and avoid overage fees (e.g., Onavo, WatchDogPro, DataWiz) \cite{datawiz}. And such trends are not exclusive to the U.S.; in South Africa, for instance, consumers' pricing plans for wired networks generally include monthly data caps. To avoid overage charges, many use ISP-provided usage-tracking tools \cite{chetty2012you}. Even in the U.S., research on in-home Internet usage has shown that many users are concerned about their wired Internet bills and would welcome applications for tracking their data usage and even controlling bandwidth rates on in-home wired networks \cite{chetty2011my}. In fact, while they are not as widely known as similar mobile applications, some apps are offered for wired platforms that allow users to do exactly that \cite{netlimiter}. 

\subsection{Content Providers' Worries}

As large U.S. ISPs like AT\&T, Comcast, and Verizon move to eliminate flat rate plans in favor of tiered data plans with usage-based overage fees in both wired and wireless networks, these changes have triggered renewed debate on the Internet's net-neutrality and openness.  These discussions have centered around who should pay the price of congestion (i.e., content providers or consumers) and how such pricing schemes should be implemented (i.e., time-of-day, app-based bundles, etc.).  The major concerns in enabling such pricing practices arise from the possibility of paid prioritization of certain content providers' traffic, price discrimination across consumers, and promoting anti-competitive behavior in bundled offerings of access plus content.  

While such developments can indeed hurt the Internet ecosystem, one aspect that should receive more attention from researchers is the threat to data usage even under current data plans.  With Internet users becoming more cautious about the increase in their monthly bills \cite{BostonGlobe}, content providers are providing new options to downgrade the quality of experience (QoE) for their users to help them save money.  For example, Netflix has started allowing ``\emph{users to dial down the quality of streaming videos to avoid hitting bandwidth caps}''  \cite{Newman}. Additionally, it is ``\emph{giving its iPhone customers the option of turning off cellular access to Netflix completely and instead rely on old-fashioned Wi-Fi to deliver their movies and TV shows}'' \cite{Fitchard}.  Thus, the ecosystem is being driven by an attitude of penalizing demand and lessening consumption through content quality degradation. 

Researchers have started to investigate these issues broadly along two lines of work: opportunistic content caching, forwarding, and scheduling; and budget-aware online video adaptation. Opportunistic content delivery involves the smart utilization of unused resources to deliver higher QoE; for example, to alleviate the high cost of bulk data transfers, Marcon et al. \citeyear{marcon2012netex} proposed utilizing excess bandwidth (e.g., at times of low network traffic) to transmit low-priority data. Since this data transmission does not require additional investment from ISPs, they can offer this service at a discount, relieving data transfer costs for clients. While utilizing excess bandwidth introduces some technical issues (e.g., the potential for resource fluctuations), a prototype implementation has shown that they are not insurmountable \cite{laoutaris2011inter}.  Hosanagar et al. \citeyear{hosanagar2005pricing} examined the pricing of caching in content delivery networks (CDNs), finding that a profit-maximizing operator should offer two vertical classes of service--premium, for which there is some charge, and a free best-effort traffic class. Economic models have also been developed for altruistic cloud providers, i.e., those whose main concern is to deliver high-quality service at low cost in response to user queries, instead of maximizing profit \cite{dash2009economic}. Simulations of user queries that mimic those for a scientific database showed that users are able to trade off cost for faster response times.  

In a similar spirit, recent works on online video adaptation systems, such as Quota Aware Video Adaptation (QAVA), have focused on sustaining a user's QoE over time by predicting her usage behavior and leveraging the compressibility of videos to keep the user within the available data quota or her monthly budget \cite{Jiasi}.  The basic idea here is that the video quality can be degraded by non-noticeable amounts from the beginning of a billing cycle based on the user's predicted usage so as to avoid a sudden drop in QoE due to throttling or overage penalties when the monthly quota is exceeded. A common feature of these new research directions is that they address the concerns of consumers and content providers by accounting for both technological and economic factors.

\section{Pricing Data Networks}\label{sec:pricing_data_networks}

In this section, we first discuss the role that access pricing in data networks can play in achieving different functionality goals and in balancing trade-offs between conflicting goals. We then provide a review of some fundamental concepts and terminology used in the network economics literature to help readers understand the discussions on different pricing proposals in the later sections of this paper. This is followed by an overview of the important challenges and open questions in pricing research.

\subsection{Network Features} 

Data networks provide connectivity among users and ubiquitous access to content, which require them to meet a range of functionality goals such as efficiency, fairness, reliability, manageability, complexity, and implementability. But these goals can often conflict to various degrees. For example, the trade-off between fairness and efficiency in allocating bandwidth to users sharing distinct subsets of links in a data network has been addressed in recent works on multi-resource allocation \cite{joe2012multi}. Pricing as a mechanism to enable a proportionally fair allocation of bandwidth among network users has also received some attention \cite{Yaiche,Kelly-PFP}, but more research is needed to understand how pricing can be used a mechanism to achieve a desirable trade-off between fairness and efficiency.  Similarly, efficiency, fairness, and reliability goals can be hard to achieve simultaneously, although pricing-based solutions can play a role in balancing these requirements. For instance, Wagner et al. \citeyear{Wagner} developed a Nash Bargaining based congestion pricing mechanism in a military network setting to distribute jobs across multiple computing clusters for fair load-balancing while using redundancy in job replication to provide reliability.  The goals of manageability, complexity, and implementability can also conflict; measures to make network traffic manageable usually introduce some system complexity and implementation challenges. Auction-based resource allocation, such as a `Smart Market' \cite{Varian}, and congestion pricing for bandwidth allocation \cite{Murphy-Murphy,ha2012tube} both improve efficiency and manageability, but require additional client-side implementation and increase billing complexity for the service provider. Pricing may also be used as a form of power control; Saraydar et al. \citeyear{saraydar2002efficient} proposed a game-theoretic framework in which wireless users compete for power so as to maximize their quality of service (QoS). In summary, various forms of pricing have been proposed by researchers to attain some level of a trade-off between one or more conflicting goals in data networks.

The role that pricing can play in networked systems is not just limited to the Internet.  Other utilities like electricity and transportation networks have used pricing mechanisms over the years to regulate demand, such as varying off-peak and peak hour prices or tolls for network access. But access pricing until recently remained largely a topic of theoretical interest in the context of the Internet, and ISPs are only now actively exploring pricing options to manage demand.  In fact, several features of data networks differentiate its pricing paradigm from that of other networks.  First, unlike traditional electricity grids which involve power generation and its delivery in a point-to-multi-point manner, the Internet is much more decentralized and allows for a bidirectional flow of traffic (i.e., uploads and downloads).  As applications and services that require bulk data transfer from the client to the cloud (e.g., machine-to-machine communications, file backup, etc.) gain popularity, pricing can be used as a signal for regulating not only the demand for content but also the inflow of content from the edge to the core of the network.  Second, the ease of two-way communication between an ISP and its customers (due to well-designed user interfaces for most network-enabled devices) creates the potential for pricing innovations that involve real-time user reaction to feedback signals from the network. In contrast, most household devices connected to the electricity network do not have usage monitoring capabilities and have to be controlled and scheduled by a centralized controller, as in smart grids \cite{joe2012optimized}. Third, pricing can exploit the time elasticity of demand more efficiently in the case of Internet services, as many online applications (e.g., movie downloads, peer-to-peer traffic, software updates) are elastic and can be completed without user intervention in small chunks whenever the prices are low \cite{ha2012tube}.  Fourth, consumer billing, privacy, and security issues associated with Internet access pricing, especially for mobile users, are much more complex and sensitive than for electricity and transportation networks. Thus, while pricing can be useful in improving the fairness, efficiency, and manageability of resources, networking researchers also need to account for the additional complexity introduced by the above considerations.

\subsection{Network Economics Fundamentals}

In this section, we introduce some of the microeconomic theory concepts \cite{microeconomics} commonly used in the broadband pricing literature. We provide a simple example to illustrate several key ideas and terminology. Consider a last-mile Internet service provider that wishes to build a single access link serving a fixed customer base of $N$ users. The ISP will charge each user based on the amount of bandwidth reserved by that user on this shared access link. Thus, the ISP wishes to determine 1) the optimal capacity $C$ needed for the access link, and 2) the usage-based price $p$ that should be charged for each unit of bandwidth reserved on this link. 

\begin{figure}
\centering
\includegraphics[width = 0.95\textwidth]{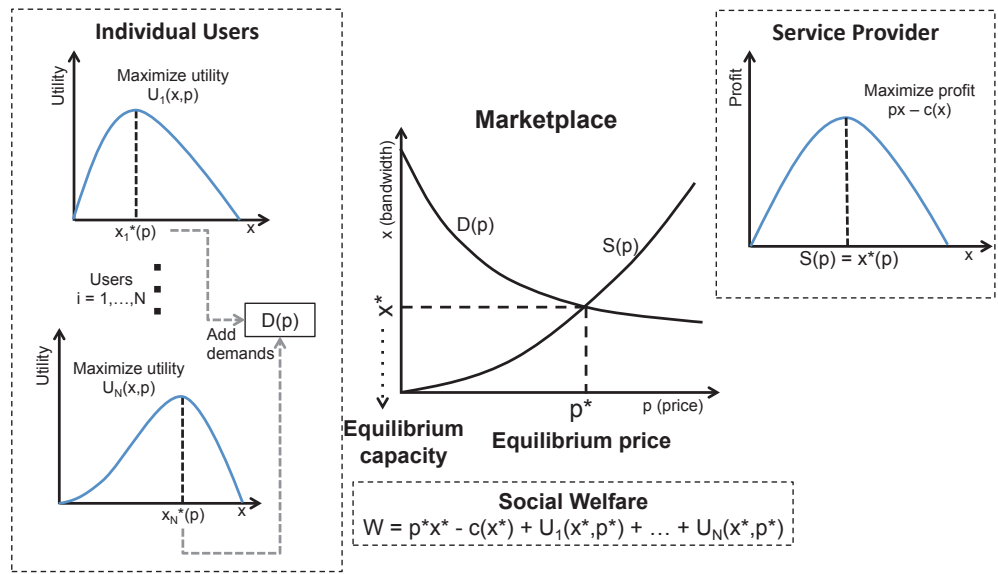}
\caption{Computation and interaction of user demand and ISP supply functions in the simplified marketplace of a single access link.}
\label{fig:supply_demand}
\end{figure}

In the traditional parlance of microeconomic theory, the ISP is a \emph{seller} and the users are \emph{buyers} of a particular \emph{good} (i.e., bandwidth).  We first consider the individual user's perspective in illustrating the economic model. We index the users by the variable $i$, $i = 1,2,\ldots, N$, and denote the bandwidth reserved by user $i$ as $x_i$. Each user is assumed to have a \emph{utility function} $U_i(x_i,p)$ that represents the satisfaction that a user receives from $x_i$ amount of bandwidth, after paying a rate $p$ for unit bandwidth. Given a price $p$ per unit bandwidth, the user can choose the amount of bandwidth used ($x_i$) so as to maximize $U_i(x_i,p)$; we then obtain a function $x^\star_i(p)$, which gives the user's optimal bandwidth usage as a function of the price. This function $x^\star_i(p)$ is then called the user's \emph{demand function}. We sum the demand functions across all users to define the aggregate demand function, $D(p) = \sum_{i = 1}^N x^\star_i(p)$, which is the total bandwidth demand as a function of the price. This utility optimization and a schematic demand function $D(p)$ are illustrated in the left side of Fig. \ref{fig:supply_demand}.

Next, we consider the ISP's perspective. We use $c(x)$ to denote the cost of building a link with total bandwidth capacity $x$. Given a price $p$ and assuming full utilization of the link, the ISP's utility or \emph{profit function} is then equal to $\Pi(x,p)= px - c(x)$, i.e., revenue $px$ less the incurred cost. The ISP can then choose the capacity $x^\star$ of the network so as to maximize its profit. Since this optimal capacity depends on the price, we denote this function as $S(p)$, i.e., the \emph{supply-side function}. The ISP's profit optimization is illustrated on the right side of the diagram in Fig. \ref{fig:supply_demand}. The individual users and the ISP then interact in the marketplace to determine the capacity $C$ of the single-link network and the price $p$. In the center of Fig. \ref{fig:supply_demand}, this interaction is represented by showing the supply function $S(p)$ and demand function $D(p)$ on the same axes. The \emph{equilibrium} operating point then occurs when supply equals demand, i.e., $S(p)$ and $D(p)$ intersect. At this price $p^\star$, each user will demand and receive her utility-maximizing bandwidth usage, and the total bandwidth $x^\star = C$ available in the network will be fully utilized.  The \emph{social welfare}, $W$, from this process is calculated by summing up over the utilities of all users and the ISP, i.e., $W=\sum_i U_i(x^\star_i,p^\star)+ \Pi(x^\star_i,p^\star)$.

\subsection{Pricing as a Congestion Management Tool}

The essential goal of smart data pricing is to incentivize users to adjust their behavior so as to better allocate ISP resources among the users. Better resource management in turn leads to a decrease in ISP cost and an increase in profit. Yet using pricing as a congestion management tool requires ISPs to think carefully about users' responses to the prices offered. In particular, the question of timescale often arises when designing pricing plans--should the prices change as network traffic volume changes? And if so, how often and by what amount?

\emph{Static pricing} changes prices on a relatively long timescale, e.g., months or years; that is, the offered prices do not vary with the network congestion level. The popularity of these plans arises from the certainty they provide regarding users' monthly bills. For instance, tiered data plans with pre-specified rates are prevalent in the United States \cite{ATT-usagebased,Verizon-plan}, while usage-based pricing in which users are charged in proportion to their usage volume is practiced by several European and Asian ISPs \cite{O2,AsiaPricing}. But even usage-based pricing leaves a timescale mismatch: ISP revenue is based on monthly usage, but peak-hour congestion dominates its cost structure (e.g., network provisioning costs increase with the peak-hour traffic). Another well-known pricing plan is time-of-day (ToD) pricing, in which users are charged higher prices for usage during certain ``peak'' hours of the day, in order to relieve congestion at these times \cite{PKF}. However, even with ToD pricing the hours deemed as ``peak'' are fixed, which results in two challenges. First, traffic peaks arise in different parts of the networks today, which
may end up creating two peaks during the day--one during peak periods, for traffic that cannot wait for several hours for lower-price periods, and another peak during discounted ``off-peak'' periods for time-insensitive traffic \cite{Economist}.  We discuss several of these existing static pricing plans and proposals in greater detail in Section \ref{sec:static}.

\emph{Dynamic pricing} goes further than ToD pricing and does not pre-classify peak and off-peak periods, instead adjusting prices at a finer timescale in response to the network congestion level. For instance, in MacKie-Mason and Varian's auction-like ``Smart Market'' approach, users ``bid'' on packets, with the bids reflecting their willingness to pay to send their packets through the network at the given time; the gateway then admits packets with the highest bids so as to maintain network efficiency and charges users according to the minimum successful bid \cite{Varian}. Congestion-dependent pricing adapts the price charged in response to the observed congestion on the network, with higher prices corresponding to higher congestion levels \cite{Kelly-PFP,Paschalidis}. However, prices that fluctuate depending on the current network load may be inconvenient for users.  Hence, another variant on dynamic ToD pricing, known as the day-ahead pricing, has been proposed to guarantee the prices a day in advance to give users some certainty about the future prices on offer. Each day, new prices are computed for different times (e.g., hours) of the next day, based on predicted congestion levels \cite{ha2012tube,Carlee}.  A detailed discussion on these dynamic pricing proposals will be provided in Section \ref{sec:dynamic}.

\begin{figure}
\centering
\includegraphics[width = 0.98\textwidth]{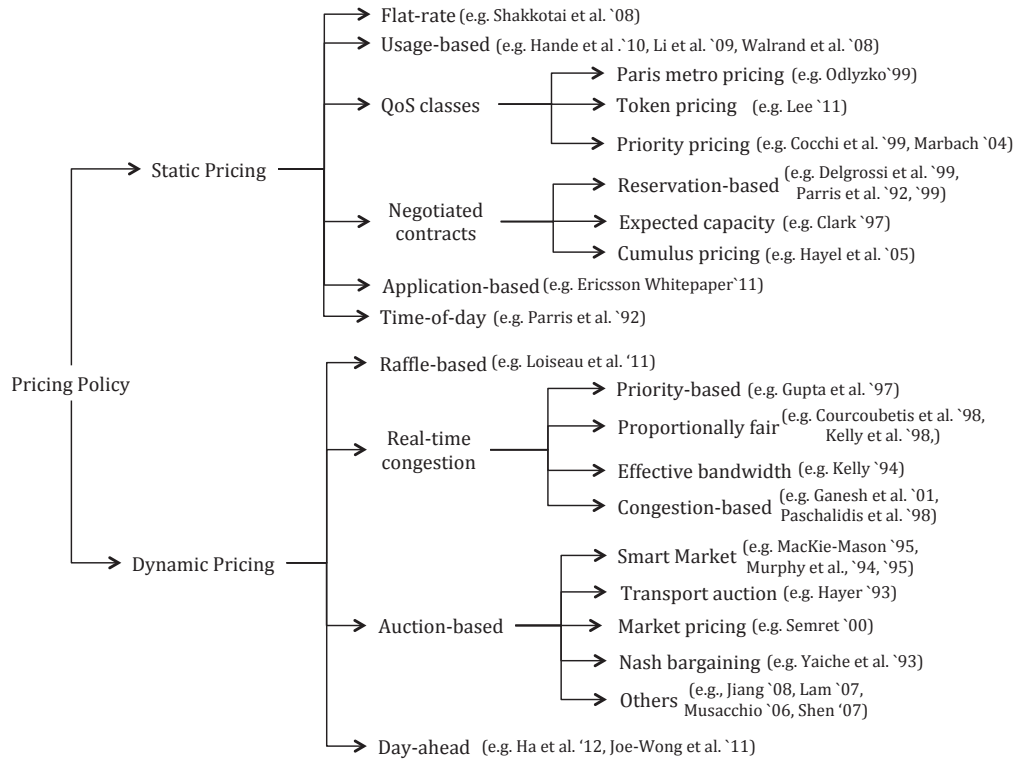}
\caption{Taxonomy of selected pricing strategies and some related proposals.}
\label{fig:taxonomy}
\end{figure}





 
The above discussion raises several fundamental questions on the timescale of broadband data pricing--should prices be based on real-time congestion? Should these price computations be done online or offline for ensuring system scalability? While real-time and other forms of dynamic pricing allow ISPs to quickly react to congestion and manage their network resources, rapidly fluctuating prices may not be convenient for consumers, requiring an automated system for submitting bids or reacting to price changes on the client-side and a scalable monitoring and price computation system on the ISP-side. These and other research challenges are discussed in the next section.

Figure \ref{fig:taxonomy} summarizes the pricing plans covered in this survey paper, including the static and dynamic plans mentioned above. Before discussing the details of these plans, however, we turn to the research challenges posed by broadband data pricing.

\section{Research Challenges in Access Pricing}\label{sec:challenges}

For more than a decade, researchers have proposed several static and dynamic access pricing plans for data networks (discussed in Sections \ref{sec:static} amd \ref{sec:dynamic}). While most of these have remained as models of theoretical interest till 2010, the growing problem of network congestion has created the possibility for realizing new pricing models and schemes in practice. This evolving landscape of access pricing is driven by moves from both wired and wireless ISPs to innovate their data plans as well as by recognition from regulators of the need for such a policy change. Consequently, this renewed interest in pricing has ushered in a new wave of research on \emph{Smart Data Pricing}  \cite{SDP}, as evidenced by the growing number of conference tracks and workshops dedicated to this topic \cite{SDP,WPIN,NetEcon}.  But research in access pricing is not just about revisiting old ideas, because the Internet ecosystem has evolved rapidly over the last decade. In particular, pricing research today has to account for several factors that have emerged in recent times, such as: (a) the inherent time elasticity of demand of several mobile applications, (b) growth in the traffic volume uploaded from the edge devices to the network core, (c) the need for better system scalability and privacy protection in a highly mobile environment, and (d) the ubiquitousness of smart devices with improved graphical user interfaces that enable faster reaction from consumers, thus paving the way to realize demand-response models and other dynamic pricing schemes.

Another key difference is that in addition to appreciating the existing pricing works, researchers need to account for both technological challenges as well as socioeconomic factors which determine the eventual adoptability and implementability of such pricing plans. Current research on Smart Data Pricing therefore focuses on verifying the implementability and efficacy of the proposed models through consumer surveys and field trials of the pricing solutions \cite{ha2012tube,sigchi,dyaberi2012managing}. Future research will likely see more such holistic and \emph{multi-disciplinary} undertakings between researchers and professionals from engineering, network economics, information systems management, human-computer interaction, marketing and industry \cite{SDP}. In the next section, we outline several design challenges, open questions, and socioeconomic challenges that researchers need to account for in their work.


\subsection{Technological Challenges}

\subsubsection{Price Computation}

Determining the optimal price to charge for network access requires developing economic models that account for the ISPs' cost structure (i.e., the trade-off between the cost of overprovisioning and the cost of network congestion) in computing the prices that they are ready to offer and that their consumers are willing to accept.  In doing so, the model may require estimating users' delay tolerance for different traffic classes and their deferral patterns in response to offered prices. The key challenges to this are choosing the correct form for utility functions, identifying methods to profile usage behavior, and using these to estimate users' price-delay tradeoffs, all while keeping the model computationally tractable.  Moreover, for scalability reasons, the ISP may need to perform these estimations without monitoring each individual household's demand pattern so as to avoid additional complexity and communication overhead \cite{Carlee}. 

Another consideration in price determination is the computation of price elasticity of demand. This is of particular interest in the case of dynamic pricing, which requires some implicit estimation of the users' elasticity across different traffic classes. However, the price a user is willing to pay for individual actions and the elasticity of demand for individual applications is often hidden inside the ``bundles'' that users purchase. Lastly, coupling dynamic pricing with dynamic QoS is another research area that requires considerable exploration.
 
\subsubsection{Communication with Users}

The mechanism used to inform users about their usage and available prices also gives rise to several non-trivial research questions: (a) finding ways for ISPs to communicate these information to the user, (b) helping users understand the impact of their usage decisions on their data quota, (c) allowing distributed control of usage across multiple devices under a shared data plan, and (d) understanding the dynamics of quota allocation among users, etc.

The first question on identifying suitable methods of communicating information to end users depends largely on how the pricing scheme works; for example, in a very dynamic pricing plan, usage decisions will need to be fully automated based on some budget set by the user and auction mechanisms to determine how packets are prioritized \cite{Varian}, while in less dynamic settings, it may be sufficient to use an application interface for communication \cite{ha2012tube}. For the second question, developing usage prediction algorithms and integrating these in data monitoring software applications \cite{datawiz} can help users to better understand the possible usage implications of their actions.  As for the latter two questions, researchers have only begun to investigate the full implications of pricing \cite{WITS} and quota dynamics of the recently deployed shared data plans.  In particular, user trials to understand how pricing practices can affect usage behavior and group dynamics among shared data plan users are some interesting directions for future networking research.


\subsubsection{Scalability and Functionality Separation}

\begin{figure}[t]
\centering
\includegraphics[width = 0.7\textwidth]{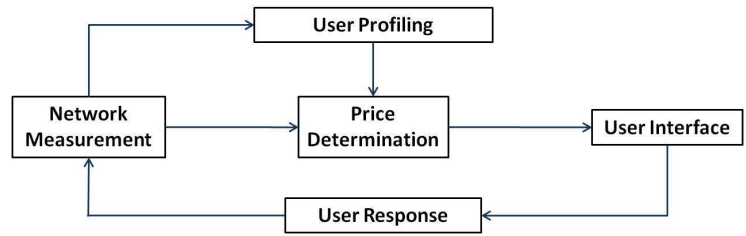}
\caption{Schematic of a feedback-control loop between users and ISPs.}
\label{fig:feedback-loop}
\end{figure}

One of the design goals of pricing solutions should be system scalability, which requires the identification and separation of client-side and server-side functionalities.  For example, dynamic pricing for data needs a feedback-control loop between the ISP server, which computes the prices to offer to users, and the users who respond to these prices \cite{MacKie-Mason}.  The ISP-side functionality will typically consist of near real-time traffic monitoring, estimating user delay tolerances for various traffic classes, and computing the prices to offer based on aggregate traffic measurements.  On the user side, functionalities like the ability to view future prices and usage history can be provided by installing applications on users' mobile devices.  Figure \ref{fig:feedback-loop} provides a schematic showing the components of this loop. 

Ensuring pricing scalability also requires developing models and algorithms that allow ISPs to compute future prices based on aggregate congestion levels, rather than monitoring each individual customer's usage and response.  Other open research questions involve the identification of mechanisms to manage shared data caps in a scalable manner.

\subsubsection{User Personalization}

The personalization of pricing plans has recently received some attention from ISPs, such as in app-based pricing and app bundles, discussed in Section \ref{sec:app-pricing}.  The key challenge in dealing with user personalization of data plans is that different users generate different traffic usage patterns by application, location, and time. Consequently, identifying the factors on which such plans should be based and measuring these plans' efficiency need more investigation.  Moreover, researchers also need to address the issue of how to track and monetize app-based transaction charges in case of such personalized plans.

In addition to the wide variation in app usage among different users, app-based pricing faces significant technological challenges on its own. For instance, without deep packet inspection software, ISPs may find it difficult to identify the traffic corresponding to different apps. Different users, in fact, may at any given time be using different versions of the same app, which can impact the network in different ways. Moreover, some traffic may not be easily classified as belonging to specific apps, e.g., clicking on hyperlinks in certain apps may open the link in an external web browser. Such ``out-of-app'' traffic may confuse the user as to whether that traffic is counted as part of the original app.

\subsubsection{Consumer Privacy and Security}

The design of access pricing solutions should also account for user privacy, particularly in terms of the feedback that ISPs record from their users and the security of the data.  For instance, if users transmit their usage statistics or some measure of willingness to pay from client devices to ISP servers, then this data exchange should take place over a secure connection.  In order to view such personal information, users might be required to sign into an account with their ISP for authentication. Such privacy issues are particularly important for pricing plans that involve per-flow pricing based on willingness to pay; these plans are highly customized to each user and thus involve a substantial exchange of personal information. A potential way to avoid privacy issues is to minimize the need to monitor individual users' usage patterns, e.g., by offering uniform prices based on aggregate network loads. Considering only aggregate usage data also helps to ensure scalability to an ISP's full customer base.  For any form of dynamic pricing, however, researchers need to address the security aspects of recording, storage, and transfer of information from client devices to ISP servers.
 
\subsubsection{Mobile Platform and Network Support}
 
Innovating access pricing requires support from software platforms on user devices and an ISP's network infrastructure, particularly for plans that aim to account for users' response to offered prices. But creating a feedback-control loop as shown in Fig. \ref{fig:feedback-loop} requires developing functionalities for client-side software. However, many widely used platforms (e.g., iOS, Android, and Windows) have different levels of platform openness and can hinder the implementation of such functionalities on the client device.  The iOS platform for iPhones and iPads has several restrictions: it strictly specifies what kind of applications can run in the background and further prevents any access other than the standard application programming interfaces (APIs). For example, obtaining an individual application's usage and running a pricing app in the background are prohibited. 
By contrast, the Android and Windows platforms allow these features, e.g., introducing an API to report individual applications' usage to third-party apps. 

Wireless ISPs' current billing systems (including 2G, 3G, and 4G) heavily depend on the RADIUS (Remote Authentication Dial In User Service) protocol, which supports centralized Authentication, Authorization, and Accounting (AAA) for users or devices to use a network service \cite{rfc2865}. In particular, RADIUS accounting \cite{rfc2886} is well suited to support usage-based pricing, since it can keep track of the usage of individual sessions belonging to each user. Interim-update messages to each session can be sent periodically to update the usage information. However, RADIUS accounting lacks support for dynamic pricing plans, which require time-of-day usage at various time scales (e.g., hourly, 30 mins or 10 mins).\footnote{Note that interim update messages are sent periodically when a session joins the system, and hence, the time interval for interim updates should be kept low to support sending time-of-day usage, which may introduce significant control overhead.} Consequently, several protocols need to be extended to support these new pricing ideas.

Another interesting direction is to create an open API between user devices and an ISP's billing systems. The open API will foster innovations in pricing for both consumers and providers. For example, the user devices connected to the ISP's network can easily fetch their pricing, billing, and usage information from the network, and the ISP can also easily test and deploy new pricing schemes through the standard interface.

\subsection{Socioeconomic Challenges}

\subsubsection{User Convenience}

Consumers typically prefer the simplicity of flat rate pricing plans \cite{Odlyzko-Public}, and hence the possibility of inciting negative public reaction is a serious concern for ISPs in experimenting with their data plans. But a combination of smart marketing and efforts to educate consumers has in the past overcome such barriers.  For similar reasons, consumers usually prefer a static over a dynamic pricing plan because of the uncertainty associated with dynamic pricing.  But this too can be overcome if researchers find the right implementation for dynamic plans and easy-to-use graphical user interfaces to communicate the price- and usage-related information.  
There are several precedents for dynamic pricing in other markets; for example, electricity markets in the U.S. have implemented both real-time and day-ahead pricing \cite{EIA}, which, with the proliferation of smart meters, are proving to be successful business models.  Similarly, in the case of road networks, congestion pricing has been accepted by the public and legislators in cities like Oslo and Rotterdam \cite{Harsman}, where residents agree with the need for tolls to decrease peak time traffic.  Moreover, the ability to collect tolls with electronic road pricing has enabled even dynamic, congestion-dependent tolls.  The successful adoption of such pricing innovations in these markets indicates that similar pricing ideas may be viable for data networks as well. 

To innovate pricing plans for data networks, researchers have proposed mechanisms that have different timescales of price adjustments. In Ha et al.'s work \citeyear{ha2012tube}, the prices vary in each hour but are computed and shown to users one day in advance through an application that runs on their mobile devices (i.e., day-ahead pricing).  In contrast, MacKie-Mason and Varian \citeyear{Varian} proposed a ``Smart Market'' scheme that requires real-time response, in which a dynamically varying per-packet price is imposed and users have to place ``bids" on each packet to send them into the network. Such an auction-based mechanism requires client-side software for fully-automated user responses, improved QoS solutions, and upgrades in the ISP's billing infrastructure. Researchers therefore need to study not only the impact of the time-scale of price adjustments on user behavior and convenience, but also the design and implementation aspects of deploying new pricing plans. 
 
\subsubsection{User Budgets}
Consumer trials have shown that mobile data customers exhibit different ``willingness to pay,'' depending on the application under consideration. For instance, Chen and Jana \citeyear{chen2013} conducted a study in which participating users experienced four different speed tiers, each about twice the speed of the previous one. Users were separated into two groups; they either received first higher and then lower, or first lower and then higher, speeds. After these experiences, they were asked how much they were willing to pay for each speed tier. 
Several users reported remarkably similar willingness to pay for different speed tiers, partly because higher speed tiers do not always result in observable gains across the whole range of applications due to the inherent heterogeneity in apps' bandwidth needs (e.g., streaming experience improves with higher data speeds but browsing activities, like mailing, stock apps, are less likely to noticeably improve). In fact, a separate trial \cite{chen2013} asked participants to use specific apps at the different speed tiers. Participants' willingness to pay for slower speeds dropped by a factor of more than two between low-bandwidth stock apps and streaming YouTube videos. Thus, depending on the applications that a person typically uses, his or her willingness to pay for different levels of service may differ significantly from that of other users. In light of this ambiguity, Chen and Jana propose a framework to allocate different speeds to different users, based on their willingness to pay as negotiated through a second-price auction.

Another determinant of users' different willingnesses to pay for different speed tiers of service is individual psychology. A recent study by Dyaberi et al. \cite{dyaberi2012managing} offered users a probabilistic monetary incentive to stop using data for 10 minute intervals throughout artificially designated ``peak hours'' of the day. The authors found that psychological measures such as ``agreeableness'' (compassionate vs. cold) and ``openness'' (curious vs. cautious) were correlated with users' acceptance of incentives, while ``neuroticism'' (nervous vs. confident) was associated with non-compliance.\footnote{The authors have three hypotheses related to psychological traits of consumers in their work. Agreeableness hypothesis: Someone who is more agreeable will generally be more compliant with requests made of them. Neuroticism hypothesis: Someone who is more likely to experience significant negative emotions will be less likely to be compliant. Openness hypothesis: Someone curious about new possibilities will be more willing to comply with a new possible mode of usage.} Interestingly, such psychological measures were found to be more significant than the type of application in this study, perhaps due to the probabilistic nature of the incentives offered. The authors did not, however, explore the resulting gains in network efficiency from an ISP's perspective, which is an interesting direction for future study.
 
\subsubsection{Ethical Concerns}

Any pricing innovation will also need to overcome arguments related to ethical issues; for example, consumer advocates have argued that dynamic pricing schemes are unfair to low-income groups, as these users are unable to reduce their usage in peak periods because they have very little usage to begin with.  On the other hand, proponents of pricing innovations have pointed out that flat-rate pricing effectively forces the subsidization of heavy users by median and low bandwidth users, and that dynamic pricing enhances users' economic efficiency \cite{sen2012incentivizing}.  This is because dynamic pricing creates the option for all consumers to choose how much they are willing to spend by deciding when and how much bandwidth they want to consume \cite{ha2012tube}.  This view is shared by economists, who describe dynamic pricing for the Internet as a natural extension of existing network control feedback mechanisms that bring ``\emph{users back into the loop and thereby ensure that performance measures are user-oriented}'' \cite{MacKie-Mason}.  Recent studies in electricity markets have further strengthened this view by confirming that 80\%-90\% of low-income households actually stand to gain from dynamic pricing of electricity \cite{faruqui2010}. Networking researchers should aware of these ethical arguments regarding access pricing in order to overcome policy hurdles. 
   
\subsection{Regulatory Challenges}

Pricing in data networks has remained a politically charged issue, particularly for pricing mechanisms that can potentially create incentives for price discrimination, non-neutrality, and other anti-competitive behavior through app-based pricing or bundling of access and content.  Academics in law schools have already cautioned that the ongoing debate on network neutrality in the U.S. often overlooks service providers' need for flexibility in exploring different pricing regimes \cite{Yoo}:   
\begin{quote}
\emph{Restricting network providers' ability to experiment with different protocols may also reduce innovation by foreclosing applications and content that depend on a different network architecture and by dampening the price signals needed to stimulate investment in new applications and content.}
\end{quote}

But faced with the growing problem of network congestion, there has been a monumental shift in the regulatory perspective in the US and other parts of the world.  This sentiment was highlighted in FCC Chairman J. Genachowski's 1 December 2010 statement \cite{FCC}, which recognizes \emph{``the importance of business innovation to promote network investment and efficient use of networks, including measures to match price to cost."}  


\section{Static Pricing}\label{sec:static} 

\subsection{Fixed Flat-Rate Pricing}

\subsubsection{Monthly Rate}

In the past, ISPs have charged users a flat monthly fee for broadband access, irrespective of the actual time spent on the network or data usage. Such a simple pricing model was praised by Anania and Solomon \citeyear{anania} and Odlyzko \citeyear{andrew} based on historical and political precedence.  A more theoretical argument was made by Shakkotai et al. \citeyear{shakkotai}, showing that simple flat-rate pricing is quite efficient in extracting revenue for elastic traffic if all users run an identical application, but with different valuations for the application. 

There are several variations of the monthly flat-rate pricing model in the real world.  For example, ``unlimited" data plans put no cap on the bandwidth used every month. If a maximum usage limit is predetermined according to a flat price, the plan is called ``flat up to a cap.''  Exceeding the limit usually incurs penalty costs that are proportional to the usage above the cap, i.e., ``metered'' pricing. ``Tiered" data plans with different flat-rate pricing for different usage caps are often used to provide a range of choices to consumers.  
Usually, users on flat-up-to-a-cap plans are shifted to usage-based pricing or a higher-priced tier upon exceeding their data limit.  But another emerging trend is ``flat to a cap, then throttle,'' as offered by Orange Spain and Comcast \cite{Spain,Comcast-cap}.



Without such penalties, flat-rate plans will simply become unviable for ISPs.  As if to drive home that point, TelstraClear, New Zealand's second largest telecommunications provider, experimented with a free weekend plan during which they switched off their usage metering and removed data caps from Friday evening to Sunday midnight of 2-5 December 2011. When all their data hungry customers simultaneously descended on the net with whetted appetite for the big weekend buffet, many were dismayed to find their speeds down to one-fifth of the usual \cite{TelstraClear}. 


\subsubsection{Hourly Rate}

Some providers have begun offering flat-rate plans in which mobile internet services are billed by the hour, i.e., the cap is specified in terms of time instead of usage. For example, USB modem customers of the Egyptian mobile operator Mobinil can choose from packages of 30 hours for EGP 80/month or 60 hours for EGP 125/month \cite{Mobinil}. 
Time-based pricing has also been proposed for road networks, in which users are charged based on how long it takes to travel from one location to another \cite{may2000effects}. However, a London study showed that time-based pricing encourages reckless driving by rewarding users who drive faster \cite{richards1996london}.

Even in broadband networks, hourly and other flat-rate pricing plans have significant disadvantages. Although flat-rate billing is cheap to implement and operate, encourages user demand, and creates simple and predictable monthly fees for customers, it suffers from several disadvantages. First, flat-rate plans lead to inefficient resource allocation and market segmentation, with low-usage customers typically subsidizing the heavy users (i.e., the bandwidth hogs) \cite{Hendershott}. Second, while ISP revenues depend on the median user, peak load costs are driven by the heavy users, which creates a price-cost mismatch. In the Berkeley INDEX project, Varaiya concurred with many of these disadvantages \citeyear{varaiya1999pricing}. 
Consequently, as the demand for bandwidth grows, most ISPs are replacing their flat-rate plans with usage-based or ``metered" data plans.

\subsection{Usage-Based Pricing}

New Zealand conducted an early experiment with usage-based pricing in 1989, when six participating universities agreed to pay by the volume of traffic sent in either direction through the NZGate gateway \cite{nevil}.  Such a pricing plan was necessitated by New Zealand's geographically isolated location, expensive trans-Pacific links, and the lack of subsidies from the New Zealand government \cite{McKnight}.  This arrangement provided an early demonstration that metering traffic and charging users by daily volume is feasible.  

``Metered" pricing implies that a user is charged in proportion to his or her actual volume of data usage \cite{walrand,Li}. In practice, operators often use ``cap then metered" plans (also known as ``usage-based" pricing), for which a user pays a flat price up to a predetermined volume of traffic, beyond which the user is charged in proportion to the volume of data consumed \cite{hande}. These variations on usage-based pricing are the prevalent pricing plans in many countries, including the U.S., U.K., Korea, Japan, and Hong Kong \cite{ATT-mobile,O2,AsiaPricing}.


Similar pricing schemes have also been implemented for other networks. For instance, analogous \emph{distance-traveled} pricing, in which drivers are charged based on the distance they travel, has been implemented for vehicle traffic in the U.S. \cite{holguin2006comparative}, as well as for trucks in Switzerland, Austria, and Germany \cite{mckinnon2006review}.


\subsection{Paris Metro Pricing}

Paris Metro Pricing (PMP) was proposed by Odlyzko \citeyear{Odlyzko} as a simple and elegant solution for creating differentiated service classes. PMP partitions the network resources into several logical traffic classes, each of which is identical in its treatment of data packets but charges users differently.  Thus, users willing to pay more will select the more expensive, and hence less congested, logical traffic class.

PMP is designed to enable maximum simplicity for the end-user in expressing his/her preference through self-selection of the desired service level. 
But Odlyzko also identified some potential problems in implementing PMP, such as finding ways to set the prices and capacities of the logically separate channels, and to maintain predictable performance on the different channels to avoid network instability. PMP may also require better designs for the user interface to let users dynamically alter their preferences and assign application sessions to different traffic classes. Recently, Wahlmueller et al. \citeyear{wahlmueller2012pricing} proposed a more user-centric variant on PMP in which QoS classes are replaced by QoE classes, with parameters such as bandwidth rate and packet loss adjusted so that the user's quality of experience remains at a specified level for different types of applications (e.g., streaming requires higher bandwidth than web browsing to achieve the same QoE). Users can also change their desired QoE, and the corresponding price they are charged \cite{reichl2010charging}.

Despite the deployment challenges for broadband networks, PMP takes its name from an actual pricing scheme on the Paris metro that was used until the 1980s to differentiate first and second class coaches. A similar concept is used today in the U.S. for roads with high-occupancy vehicle (HOV) lanes, which are identical to all other lanes of the road, except that HOV lanes are restricted to vehicles with multiple passengers \cite{hov}.  Drivers can self-select to use less congested HOV lanes by paying the higher ``price'' of carpooling with other passengers.

\subsection{Token Pricing}

Lee et al. \citeyear{Token} introduced a token pricing scheme in which users pay a fixed flat-rate monthly fee for Internet access.  Each user receives a certain number of ``tokens" from the service provider, which offers two service classes of different qualities. The basic quality class requires no tokens for access but may become congested.  The higher quality class requires users to redeem some tokens, and hence has lesser congestion and a better service because of a PMP-like effect. Thus, users self-prioritize their sessions and implicitly pay for the higher QoS with their tokens for particularly urgent sessions during peak network congestion. Each session, regardless of size, costs the same number of tokens, and users continually receive more tokens (e.g., one every ten minutes). The benefits of such a system in reducing peak network congestion in a real network have yet to be explored. 

\subsection{Priority Pricing}

Cocchi et al. \citeyear{Cocchi1,Cocchi2} studied a pricing scheme in a multiple service network with priority classes, but without any resource reservation.  Users 
can request different QoS by setting bits in their packets.  A higher priority class charges a higher per-byte fee but is assumed to receive better service from the network. Thus, users who pay a greater per-byte fee for higher priority are in effect paying for the negative externality imposed on traffic from other, lower priority users.  The authors showed that such quality-sensitive pricing is more efficient (in a Pareto sense) than flat pricing. However, this result depends on the reservation-less assumption.



 A game-theoretic framework was used to analyze static priority pricing by Marbach \citeyear{Marbach}.  In his single-link model, users assign a priority class to their packets and are charged accordingly. These charges are based on the packets submitted to the network rather than their actual delivery. Marbach showed that there always exists a Wardrop equilibrium bandwidth allocation, but it is not necessarily unique. This equilibrium allocation and the associated link revenue do not depend on the prices of the different priority classes, but the prices do have a simple relation with the packet loss probability due to congestion in that class.  

One disadvantage of priority pricing is users' inability to express their desired levels of delay and bandwidth share; priority pricing's consistent preferential treatment of higher priority classes might drive lower priority classes to little or no usage.  Priority pricing is largely absent today, but some ISPs are considering the idea of creating a ``priority data plan'' in which users of the premium service are prioritized during periods of network congestion. Recently, SingTel of Singapore introduced such an option, called ``priority pass,'' for its top-tier dongle customers \cite{SingTel}.

\subsection{Reservation-Based Pricing}

Parris et al. \citeyear{PKF} were one of the first to study pricing in a reservation-oriented network. They considered the issues of network utilization, ISP revenue, and blocking probability under per-packet, setup, and peak-load pricing schemes. In their work, users are characterized by their connection durations, budgets, and chosen classes of service (with a higher per-byte fee for a higher priority service class). The network decides to either accept or block the connection, depending on the sufficiency of the user's budget and the availability of network resources. Using simulations, the authors show that for a given pricing scheme, price increases will at first increase and then eventually decrease the net revenue, but will always decrease the blocking probability and network utilization. Setup pricing decreases the blocking probability and increases revenue for the ISP, and more generally performs better than per-packet pricing (in that the blocking probability from admission control is lower under setup pricing than per-packet pricing for the same level of revenue generated).  

Despite these advantages, Parris et al.'s \citeyear{PKF} form of reservation pricing suffers from some shortcomings. First, a flat-rate setup cost is unfair towards those users with shorter conversations. Second, poorer users may not be able to afford a connection under a high setup cost, thus leading to a greater digital divide. Third, average network utilization is typically lower in the presence of setup costs, thus demonstrating a tradeoff between network efficiency and revenue maximization. 

Parris and Ferrari  \citeyear{ParrisF} presented another reservation pricing scheme for real-time channel establishment.  Under this plan, users are charged based on the type of service requested, as measured by factors such as the bandwidth, buffer space, CPU time resources reserved, and delay imposed on other users. The total charge a user pays is a product of the type of service measure, channel duration, and a time-of-day factor. 
However, the work does not provide any clear guidelines on how economic considerations are to be mapped to a single time-of-day factor or the impact of the overhead associated with estimating network parameters in real time.

In a later work, Delgrossi and Ferrari \citeyear{Delgrossi} considered a pricing scheme based on the portion of resource capacity used by a reserved data channel in a multiple-service network. 
They introduced a charging formula with different reservation and transport cost components for real-time and non-real-time traffic, along with a discussion on computing resource capacity requirements for the channel as a function of the buffer size, processing power, and schedulability.  

\subsection{Time-of-Day Pricing}

Time-of-day or ToD pricing schemes charge peak and off-peak hours differently, so as to disperse user demand more uniformly over time. Parris et al. \citeyear{PKF} and Parris and Ferrari \citeyear{ParrisF} considered a form of ToD combined with reservation-based pricing, which divides a day into peak and off-peak periods and incorporates the time elasticity of user demand. They show that peak-load pricing reduces the peak utilization and the blocking probability of all traffic classes, and increases revenue by inducing a more even distribution of demand over peak and off-peak periods.  

The most basic form of ToD in practice is a two-period plan that charges different rates during the daytime and night time.  For example, BSNL in India offers unlimited night time (2-8 am) downloads on monthly data plans of Rs 500 (\$10) and above. Other variations of ToD pricing are offered elsewhere; for instance, the European operator Orange has a ``Dolphin Plan''  for $\pounds$15 (\$23.50 USD) per month that allows unlimited web access during a ``happy hour'' corresponding to users' morning commute (8-9 am), lunch break (12-1 pm), late afternoon break (4-5 pm), or late night (10-11 pm).


Time-of-day pricing has been implemented in road networks since the 1990s, e.g., in California, Norway, and Sweden \cite{RoadPricing}. The electricity market has also practiced static, ToD pricing for many years, a move accelerated by power shortages over the past decade.  Many academic works on ToD pricing in these markets have also focused on constructing user demand functions for ToD pricing by using data on electricity consumption, focusing on the case of two prices (peak and off-peak) \cite{faruqui2008quantifying,hausmann1979two}. Several works have reported on consumer trials of time-of-day pricing \cite{CharlesRiver,faruqui2009piloting,herter2007residential,matsukawa2001household,board2007ontario,wells2004electricity,wolak2006residential}.  Many of these trials employed a variant on ToD, called \emph{critical peak pricing}, in which the price on certain days of especially high electricity consumption (e.g., very hot days during the summer) was increased beyond the normal peak price. Such pricing was generally judged to be successful in lowering peak electricity consumption, especially on days of high demand.

\subsection{Expected Capacity Pricing}

In 1997, Clark \citeyear{DDClark} wrote 
\begin{quote}
\emph{In the future it will be desirable to provide additional explicit mechanisms to allow users to specify different service needs, with the presumption that they will be differentially priced}.
\end{quote}
He proposed expected capacity pricing as a mechanism to allow users to explicitly specify their service expectation (e.g., file transfer time), while accounting for differences in applications' data volume and delay tolerance. 
The idea is that by entering into profile contracts for expected capacity with the operator, different users should receive different shares of network resources \emph{only} at times of congestion \cite{Songhurst}.  

 
One specific proposal to realize this service involved traffic flagging (i.e., each packet is marked as being \emph{in} or \emph{out} of the user's purchased profile, irrespective of network congestion level) by a traffic meter at access points where the user's traffic enters the network. This is followed by congestion management at the switches and routers where packets marked as \emph{out} are preferentially dropped during congested periods, but are treated in an equal best-effort manner at all other times.  The expected capacity is thus not a capacity guarantee from the network to the user, but rather a notion of the capacity that a user expects to be available and a set of mechanisms that allow him or her to obtain a different share of the resource at congested times.  This pricing can be simply enforced at the router and switches of the network, and allows service providers to have more stable estimates of the future necessary capacity based on the total expected capacity sold, rather than the sum of peak rates of all users' access links.  A dynamic pricing version of the scheme has also been explored \cite{DDClark}. However, in order to implement this pricing scheme, the assignment of price values to expected capacity profiles requires further study.

\subsection{Cumulus Pricing}

Cumulus pricing schemes consist of three stages: specification, monitoring, and negotiation. A service provider initially offers a flat-rate contract to the user for a specified period based on the user's estimate of resource requirements. During this time the provider monitors the user's actual usage and provides periodic feedback to the user (by reporting on ``cumulus points'' accumulated from their usage) to indicate whether the user has exceeded the specified resource requirements. Once the cumulative score of a user exceeds a predefined threshold, the contract is renegotiated.  Hayel and Tuffin \citeyear{Hayel} studied such a scheme and used simulated annealing to optimize the total network revenue in terms of the renegotiation threshold.  

Cumulus pricing is a simple pricing scheme that can be easily implemented at the network edge. Some ISPs have been experimenting with similar ideas. For example, Vodafone in the U.K. announced a new ``data test drive'' plan that allows customers joining any of the monthly pay contracts to have unlimited data access (including tethering, but excluding roaming) for the first three months. The data usage report is then fed back to the user to negotiate whether his or her chosen plan is appropriate. The user can choose to either continue with the existing plan, possibly incurring overage charges, or to switch to an alternative plan suggested by Vodafone \cite{VodafoneTest}.   

\subsection{Application- and Content-Based Pricing} \label{sec:app-pricing}

Several mobile service providers have been experimenting with various forms of application- or content-based pricing. These are somewhat analogous to certain types of tolls in road networks, in which drivers are charged different rates for different types of vehicles (e.g., trucks are charged at a higher rate) \cite{ushighways}.  Indeed, charging by vehicle type was mentioned as a criterion for efficient road pricing in a report by the U.K. Ministry of Transportation \cite{may1992road}.

While the currently available app-based data plans are mostly designed to attract and ``lock in'' customers by bundling content and data plans, they highlight an emerging trend of operators experimenting with pricing structures that charge (or subsidize) differently based on application type \cite{Ericsson,Gigaom}. These types of bundled services, which generally offer bundled access to music or movie streaming, are offered by Three in the U.K., Telus in Canada, and TDC in Denmark \cite{ContentPricing,Telus}. Orange UK allows users to choose which media services to bundle \cite{Swapables}.

Another form of app-based pricing that is becoming increasingly popular is ``sponsored content,'' in which operators subsidize certain types of data. For instance, Mobistar in Belgium offers ``zero-rated'' or free access to Facebook and Twitter \cite{Mobistar}. Japan's Softbank Mobile is also moving in this direction, having contracted with Ericsson to develop the infrastructure for app-based and other types of innovative data plans \cite{japan-ericsson}. While technological advances such as Deep Packet Inspection (DPI) have allowed ISPs to differentiate between the traffic of some apps and offer these sponsored content plans, app-based pricing still faces several challenges. As discussed in Section \ref{sec:pricing_data_networks}, these challenges will likely limit the deployment of app-based pricing at the present time.

\begin{figure*}
\centering
\includegraphics[width = 0.98\textwidth]{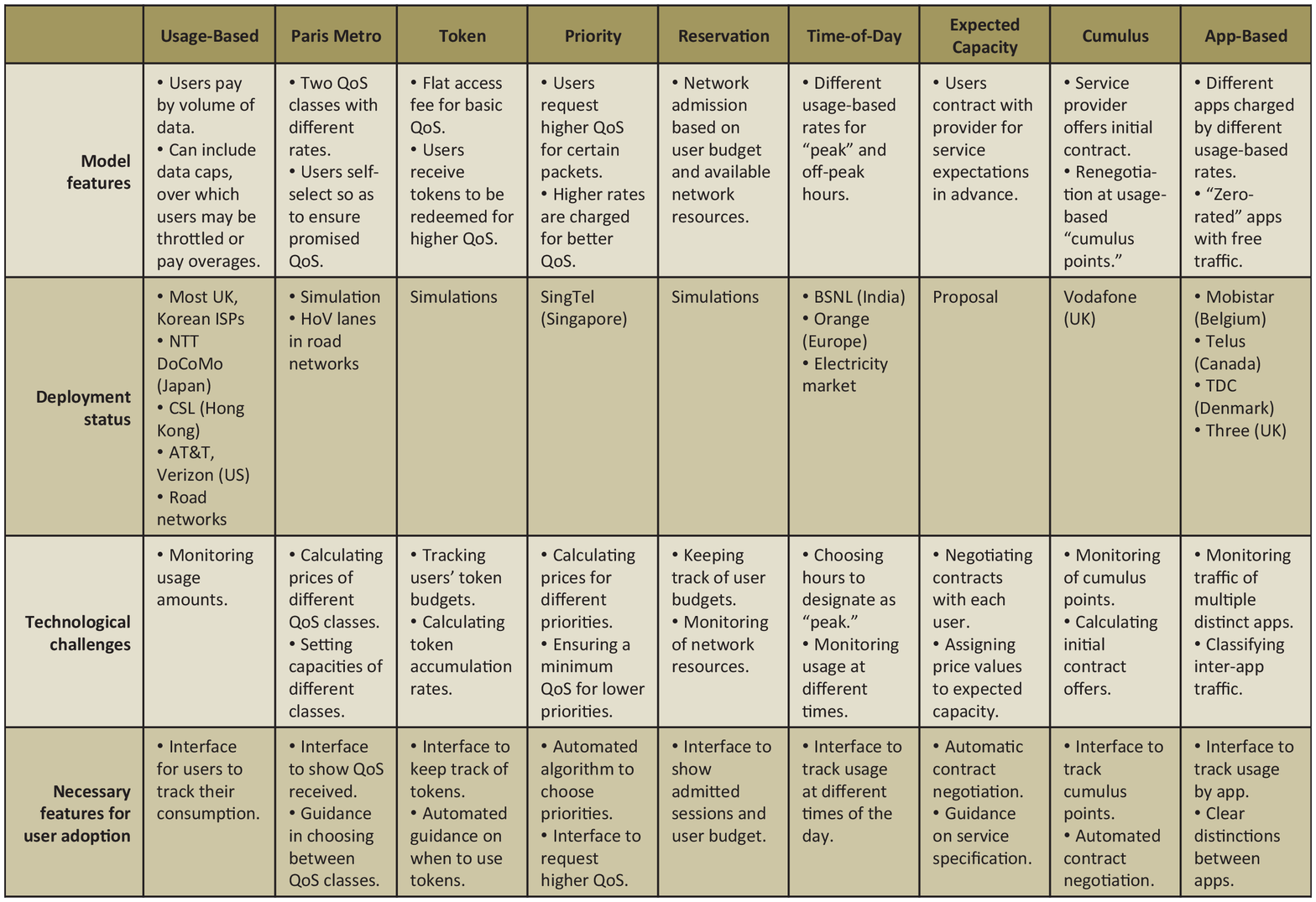}
\caption{Comparison of static pricing schemes (Section \ref{sec:static}).}
\label{fig:static_compare}
\end{figure*}

\noindent Figure \ref{fig:static_compare} compares the various static pricing schemes discussed in this section in terms of their key features, deployment status, and both technological and socioeconomic challenges. While different forms of usage-based pricing are currently the norm in most parts of the world, some operators have begun to introduce more innovative data plans, such as priority- or app-based pricing.

\section{Dynamic Pricing} \label{sec:dynamic}

\begin{figure*}
\centering
\includegraphics[width = 0.98\textwidth]{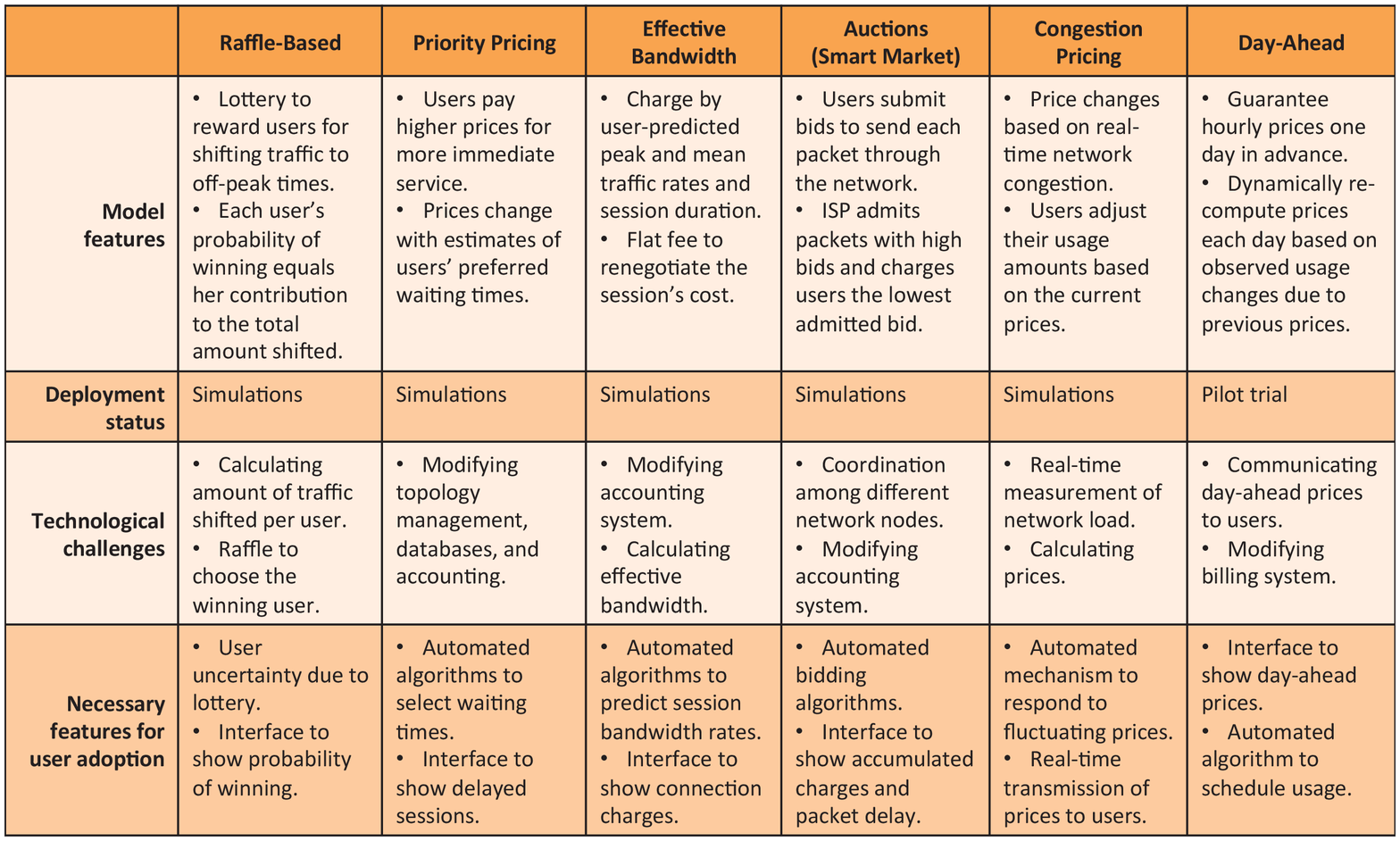}
\caption{Comparison of dynamic pricing schemes (Section \ref{sec:dynamic}).}
\label{fig:dynamic_comparison}
\end{figure*}

In this section, we discuss pricing models in which the prices offered vary over time, e.g., due to fluctuating network congestion levels. While such models often allow ``better'' network management by providing the ISP with more pricing flexibility, they also suffer from significant barriers to user adoption, as discussed in Section \ref{sec:challenges}. A summary and comparison of the pricing plans discussed may be found in Fig. \ref{fig:dynamic_comparison}.



\subsection{Raffle-Based Pricing}

Like ToD pricing, raffle-based pricing divides the day into peak and off-peak periods.  Users are then offered probabilistic incentives to shift their demand to off-peak periods, in the form of a raffle or lottery for a given monetary reward.  The probability of winning the lottery is proportional to the user's contribution to the reduction in peak demand.  This Òall-or-nothingÓ lottery may instead be replaced by one in which the ISP divides the total reward by the total amount of traffic shifted, paying this amount to each user with a probability equal to the percentage of usage shifted to the off-peak period.  Loiseau et al. \citeyear{loiseau2011incentive} derived expressions for the Nash equilibrium of this user-ISP interaction and show that in some cases the social welfare is more robust to price variations than a comparable time-of-day pricing plan with two periods.  Yet the uncertainty inherent in raffle-based pricing may reduce its effectiveness; users may not shift their usage to off-peak periods without guaranteed rewards. Indeed, the reward amount depends on other users' behavior. A field trial of such a policy would therefore be necessary to understand its effectiveness.

A recent trial by Dyaberi et al. \citeyear{dyaberi2012managing} employed a similar probabilistic incentive mechanism, offering (with some probability) rewards to users for terminating their cellular usage during a 10-minute interval. Though the authors observed an 83\% acceptance rate, their trial participants were largely university students, and the devices used were loaned for the purposes of the trial. Thus, their results may not extend to the general population.

\subsection{Dynamic Priority Pricing}\label{sec:dynpriority}

Gupta et al. \citeyear{Gupta} presented a dynamic priority pricing mechanism in which the prices serve as a congestion toll for network access. The authors modeled user service requests as a stochastic process and network nodes as priority queues. 
A user's incoming request has an instantaneous value for the service and a linear rate of decay for this value to capture the delay cost. Users' requests can be fulfilled using different alternatives, each with its corresponding price and waiting time.  The user trades off between the total cost of service and the cost of delay to choose an optimal alternative in a particular priority class. The principal benefit of such a scheme is that it achieves resource allocation in real time in a completely decentralized manner; moreover, users' myopic decisions lead to socially optimal choices.


Dynamic priority pricing is based on general equilibrium theory, which generalizes the notion of an ``equilibrium'' to several interacting markets, in this case the market situation at any particular point in time (i.e., supply equals demand at any given time). However, since computing general Arrow-Debreu equilibria in the volatile environment of the Internet is expensive, the authors introduced the notion of a stochastic equilibrium and derive optimal prices that support the unique stochastic equilibrium.  They also developed a decentralized real-time mechanism to compute near-optimal prices for the stochastic equilibrium: the cost of service (i.e., the price) is updated after each time period by taking a weighted average of the price in the previous period and the new optimal price, calculated from updated estimates of the waiting times for different priority classes.  The convergence properties of this dynamic pricing mechanism were demonstrated with simulations that allowed the system to adapt to changing loads on the network.  However, implementing dynamic priority pricing will require modifications to network topology management applications, databases, accounting systems, and the end-user interface.   





\subsection{Proportional Fairness Pricing}

Kelly et al. \citeyear{Kelly-PFP} proposed proportional fairness pricing as a means to allocate resources (which determine user rates) in proportion to the user's willingness to pay. The global optimization of maximizing net utility across all users, given resource capacity constraints, can be decomposed into a user and a network optimization problem. Kelly showed that there exist a price vector and a rate vector that optimize both the user and the network's optimization problems. Alternatively, if each user chooses a price per unit time according to his or her willingness to pay, and if the network allocates rates per unit price that are proportionally fair, then a system optimum is achieved when users' choices of prices and the network's rate allocation are in equilibrium.  Courcoubetis et al. \citeyear{Courcoubetis} extended this idea by replacing end-users with intelligent agents that can decide the willingness to pay on behalf of the user while maximizing the user's utility. However, this element introduces overhead in installing such agents on the users' devices or machines and adding network servers to compute the optimal rate allocation vectors at a short timescale.

\subsection{Effective Bandwidth Pricing}

\begin{figure}
\centering
\includegraphics[width = 0.4\textwidth]{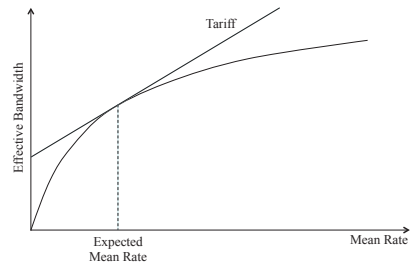}
\caption{Illustration of the tariff charged in effective bandwidth pricing.}
\label{fig:effective}
\end{figure}

Kelly's effective bandwidth pricing \citeyear{Kelly-eff} is a variant on usage-based pricing in which users are charged based on self-reported peak and mean traffic rates, as well as the observed mean rate and duration of each connection.  Before a user's connection is accepted, the user is required to provide mean and peak rates for the connection.  Given a formula describing effective bandwidth as a function of the peak and mean rates, the user is charged a tariff given by the tangent line to this effective bandwidth formula (as a function of the mean rate) at the self-reported mean and peak rates.  Evaluating this tariff at the observed mean rate, the result is multiplied by the connection duration to give the total charge to the user.  Figure \ref{fig:effective} shows a schematic illustration of such a tariff.

Kelly showed that under this pricing scheme, users minimize their expected cost by accurately reporting the connection's mean and peak rates.  Thus, the final charge to the user consists of a term proportional to the connection duration and another term proportional to the connection volume.  Users may renegotiate the tariff for a flat fee, e.g., for highly variable traffic.

This pricing scheme can also be extended to connection acceptance control--a connection is accepted if the network's effective load, as calculated from the tariffs charged to existing connections, falls below a certain threshold value.  While effective bandwidth pricing is compatible with user incentives and fairly simple, it does require an explicit effective bandwidth formula, and it requires users to know, or at least estimate, the peak and mean rates of each connection.  Moreover, further validation is needed to understand whether the benefits of this pricing scheme justify the accounting overhead associated with charging each connection based on its duration and volume.

\subsection{Responsive Pricing}



MacKie-Mason et al. \citeyear{MacKie-Mason} described the concept for responsive pricing in the following words:
\begin{quote}
\emph{By associating a cost measure with network loading, all users can be signaled with the prices necessary to recover the cost of the current network load. Price-sensitive users--those willing and able to respond to dynamic prices--increase economic efficiency by choosing whether or not to input traffic according to their individual willingness to pay the current price.}
\end{quote}
In other words, a user's price sensitivity and time sensitivity for different applications can be exploited by networks to dynamically set prices to alleviate congestion.  This process broadly encompasses the philosophy behind different forms of dynamic time-dependent pricing. In the case of the Internet, MacKie-Mason et al. \citeyear{MacKie-Mason} argued that such a responsive pricing system is required for network efficiency. 

The network can set response prices either in a closed-loop feedback \cite{Murphy-Posner,Murphy-Murphy} or a ``Smart Market'' approach \cite{Varian}.  In a closed-loop setting, the network state, measured in terms of the buffer occupancy at the gateway, is converted to a price per packet for users' adaptive applications, which then decide how much data to transmit.  This closed-loop feedback is thus similar to Gupta et al.'s priority pricing \citeyear{Gupta}, discussed in Section \ref{sec:dynpriority}: the basic idea is that dynamic prices are set so as to incentivize users to behave in a way that enhances network efficiency.  However, due to the adaptive nature of the prices, there is a slight delay in the feedback loop: prices are set based on network conditions in the previous time period.\footnote{Similar ideas have also been adopted in the electricity industry; for instance, Vytelingum et al. \citeyear{vytelingum2010trading} developed an auction scheme that takes into account varying market capacity, with dynamic offers from electricity distributors and real-time responses from users (i.e., households) buying electricity.}

This delay in the feedback loop is tightened in the Smart Market approach proposed by MacKie-Mason and Varian \citeyear{Varian}.  In this approach, each user places a ``bid'' on each packet that reflects their willingness to pay to send the packet onto the network at a given time.  The gateway admits packets in the descending order of their bids as long as the network performance remains above a desired threshold. Users are charged according to the minimum bid on a packet admitted into the network at the time, and thus pay only for the congestion cost at the market clearing price.  While such auction schemes encourage network and economic efficiency, they require substantial changes in users' application interfaces and providers' billing systems, with additional concerns in the case of billing contention, etc.  Moreover, as discussed by Sarkar \citeyear{Sarkar}, such a pricing scheme would require extensive coordination among different nodes of the network, and is open to abuse by those controlling bottleneck facilities.  Sarkar argued that these concerns may be overcome with appropriate governmental regulation.

\subsection{Dynamic Congestion Pricing}

Dynamic congestion pricing is a particular realization of the idea of responsive pricing, in which the network announces prices based on current congestion levels and the user response to these prices is fed back into the control loop to compute new prices.  Ganesh et al. \citeyear{Ganesh} used congestion prices as a mechanism to provide both feedback and incentives to end-systems for rate adaptation in a decentralized manner; they then studied the resulting dynamics.  Paschalidis and Tsitsikilis \citeyear{Paschalidis} addressed the issue of revenue and welfare maximization for dynamic congestion pricing of customer calls by using a dynamic programming formulation. In their model, users initiate calls that differ in their service class, resource requirements, and call duration.  Based on the current congestion level, the service provider charges a fee per call, which in turn influences users' demand.  Their findings additionally corroborate the usefulness of time-of-day pricing in reducing network congestion problems.

Congestion-based pricing innovations have been adopted in many markets outside of the U.S.  In recent years, network operators in highly competitive and lucrative markets, such as those in India and Africa, have adopted innovative congestion- and location-dependent dynamic pricing for voice calls \cite{Economist,Uninor}, although not yet for mobile data plans. 
An analogous congestion-based scheme for transportation networks, in which users are charged baed on both their distance traveled and time to travel that distance, was considered in Cambridge, U.K., but was never fully implemented \cite{RoadPricing}.

Congestion pricing in wireless cellular networks has also received some attention in the literature, e.g., a survey by Al-Manthari et al. \citeyear{al2011congestion}, which divides pricing polices into two categories: admission-level and power-level pricing. The authors mentioned several challenges specific to wireless networks, e.g., interference, user mobility, and bandwidth limits, and surveyed proposed pricing policies that incorporate priority queues and QoS support.

Several works on pricing in the electricity market have adopted a congestion-based dynamic pricing model, as reviewed by Borenstein et al. \citeyear{borenstein2002dynamic} and proposed in a more recent paper by Samadi et al. \citeyear{samadi2010optimal}. Many of these focus on end user management of congestion-dependent prices, proposing algorithms to predict future prices and schedule energy usage accordingly \cite{Du,mohsenian2010optimalres,mohsenian2010optimal}. Other works have attempted to combine these two perspectives by examining a feedback loop between the user and provider \cite{roozbehani2010dynamic,li2011optimal}.

\subsection{Day-Ahead Pricing} 

Day-ahead pricing is a variant on dynamic congestion pricing in which prices vary by the time of the day (e.g., each hour) and users are told the future prices one day in advance. Joe-Wong et al. \citeyear{Carlee} proposed an algorithm for calculating these day-ahead prices that models a user's willingness to shift different types of usage to lower-price times. The model accounts for the fact that some types of traffic, e.g., email, are more delay-sensitive, while for other traffic, e.g., software updates, users may be willing to wait several hours for a lower price. As prices are offered throughout the day, the ISP monitors the change in traffic volume relative to a baseline traffic trace with constant prices. These changes are then used to optimize the prices over the next day, in a feedback loop structure similar to that proposed by MacKie-Mason et al. \citeyear{MacKie-Mason}.

A recent pilot trial in the U.S. has demonstrated the feasibility, benefits, and architecture of day-ahead pricing for mobile data. In this trial, Ha et al. \citeyear{ha2012tube} implemented a form of \emph{day-ahead} pricing for a small number of iOS users and reported that such pricing policies can in fact reduce the peak load on the network. While this form of dynamic pricing less accurately reflects the real-time network congestion, it has the advantage of increasing user certainty by guaranteeing some prices in advance and allowing users to plan ahead. In fact, Ha et al.'s trial included an automated scheduling algorithm on user devices that can schedule usage based on future prices so as to save users' money  \citeyear{ha2012tube}. A similar day-ahead pricing scheme has also been proposed for electricity smart grids \cite{joe2012optimized}.




\subsection{Game-Theoretic Pricing}

Several authors have used game-theoretic models for pricing data, some of which are briefly discussed here. Hayer \citeyear{Hayer} proposed transport auctions as a way to distribute excess capacity across users with delay tolerant traffic. A decentralized auction-based approach to the pricing of edge-allocated bandwidth, called ``market pricing,'' was explored by Semret et al. \citeyear{Semret} in a differentiated services Internet. 

Ya\"{i}che et al. \citeyear{Yaiche} introduced a cooperative game-theory framework that used Nash bargaining (i.e., a competition model whose solution satisfies certain fairness axioms such as Pareto-optimality\footnote{A Pareto-optimal allocation of resources is one for which increasing any one user's allocation will decrease another user's allocation.} \cite{nash1950bargaining}) to compute bandwidth allocations for elastic services and pricing in broadband networks. This framework provides rate allocations for users that are not only Pareto-optimal from the viewpoint of the whole system, but also consistent with game-theoretic fairness axioms.

Other researchers have used game-theoretic formulations to investigate the effect of pricing on user adoption and fair use.  Pricing models to induce participation and collaboration in a public wireless mesh network were studied by Lam et al. \citeyear{Lam}.  Shen and Basar \citeyear{Basar} investigated optimal nonlinear pricing policy design as a means of controlling network usage and generating profits for a monopolistic service provider.  Dynamic game models have also been used to determine {WiFi} access point pricing by Musacchio and Walrand \citeyear{Musacchio}, which is relevant in the context of congestion management through {WiFi} offloading. 

Jiang et al. \citeyear{Jiang} introduced a model to study the role of time preferences in network pricing. In their model, each user chooses his or her access time based on his or her preference, the congestion level, and the price charged. 
The authors show that maximization of both the social welfare and the revenue of a service provider is feasible if the provider can differentiate its prices over different users and times. However, if the prices can only be differentiated over the access times and not across users due to insufficient information, the resulting social welfare can be much less than the optimum, especially in the presence of many low-utility users. Caron and Kesidis \citeyear{caron2010incentive} took a similar approach for the electricity market; in their model, users schedule their energy usage in a cooperative game so as to minimize the total load on the network.
 
Despite these theoretical works, game-theoretic models have found little traction among real network operators so far, perhaps due to the stylized nature of the theoretical models and the challenges in estimating user utility and system parameters in the real world.\\



\noindent The different real-world examples of various static and dynamic pricing plans for the wired and wireless services discussed in this paper are summarized in Table \ref{tab:plans-summary}.  

\small
\begin{table}
\renewcommand{\arraystretch}{1.2}
\tbl{List of key example pricing plans discussed.\label{examples}}{
\centering
\begin{tabular}{lllll}
\hline
\multicolumn{2}{c}{Pricing Practice} & \multicolumn{3}{c}{Example Pricing Plan (see the paper for details)} \\
\multicolumn{1}{c}{Type} & \multicolumn{1}{c}{Category} & \multicolumn{1}{c}{Description} & \multicolumn{1}{c}{Network} & \multicolumn{1}{c}{Country} \\ \hline \hline
\multirow{24}{*}{Static} & \multirow{12}{*}{Fixed Flat-Rate} & Monthly fee; & \multirow{2}{*}{Both} & Vanishing \\ 
& & unlimited & & worldwide \\ \cline{3-5}
 & & Monthly; & \multirow{3}{*}{Wired} & U.S. (AT\&T, \\
 & & flat to a cap, & & Verizon) \\
 & & then usage-based & & \\ \cline{3-5}
& & Monthly; & \multirow{3}{*}{Wired/wireless} & {Spain (Orange)} \\ 
& & flat to a cap, & & \multirow{2}{*}{U.S. (Comcast)}\\
& & then throttle & & \\ \cline{3-5}
& & \multirow{3}{*}{Monthly; shared} & \multirow{3}{*}{Wired/wireless}  & Canada (Rogers) \\ 
& &  & & U.S. (AT\&T, \\
& & & & Verizon) \\ \cline{3-5}
 & & Hourly rate & Wireless  & Egypt (Mobinil) \\ \cline{2-5}
 & \multirow{2}{*}{Usage-Based} & \multirow{2}{*}{Cap then metered} & \multirow{2}{*}{Wired/wireless} & Worldwide \\ 
 & & & & (e.g. U.S., U.K.) \\ \cline{2-5}
 & \multirow{2}{*}{Priority Pricing} & Priority pass & \multirow{2}{*}{Wireless} & Singapore \\ 
 & & (for dongle users) & & (SingTel) \\ \cline{2-5}
 & \multirow{4}{*}{Time-of-Day} & Day-time \& & \multirow{2}{*}{Wireless} & \multirow{2}{*}{India (BSNL)} \\ 
 & & night-time rates & & \\ \cline{3-5}
 & & Users choose & \multirow{2}{*}{Wireless} & \multirow{2}{*}{U.K. (Orange)} \\ 
 & & happy hours & & \\ \cline{2-5}
 & \multirow{2}{*}{Cumulus Pricing} & Usage-based & \multirow{3}{*}{Wireless} & \multirow{3}{*}{U.K. (Vodafone)} \\ 
 & & contract & & \\ 
 & & negotiation & & \\ \cline{2-5} 
 & \multirow{2}{*}{App-Based Pricing} & Free access to & \multirow{2}{*}{Wireless} & U.K. (Orange) \\ 
 & & select apps; & & Denmark (TDC) \\ 
 & & bundling & & \\ \hline \hline
 
\multirow{6}{*}{Dynamic} & \multirow{6}{*}{Congestion-Based} & Hourly price & \multirow{2}{*}{Wireless (voice calls)} & \multirow{2}{*}{Uganda (MTN)} \\ 
& & changes & & \\ \cline{3-5}
& & Location and & \multirow{2}{*}{Wireless (voice calls)} & \multirow{2}{*}{India (Uninor)} \\ 
& & cell-load based & & \\ \cline{3-5} 
& & Time- and & \multirow{2}{*}{Wireless (data)} & \multirow{2}{*}{U.S. (pilot trial)} \\ 
& & usage-based & & \\ \hline 
\end{tabular}} 
\label{tab:plans-summary}
\end{table}
\normalsize

\section{Emerging Technology Trends} \label{sec:directions}

The recent exponential growth in data demand has catalyzed major changes within broadband pricing research and practice.  In this section, we discuss the new directions that are emerging from these recent changes.

\subsection{Satellite Broadband}

This survey focuses mainly on wired and wireless pricing, as these are the mediums most impacted by users' increasing demands for data.  Recently, however, many companies have begun to offer satellite broadband as an alternative to wired or wireless Internet access.  While satellite broadband solutions have existed for over twenty years, satellite Internet access has only recently become a popular offering for end users.  Satellite is becoming especially prevalent as a way to reach users in rural and sparsely populated regions, such as in Africa, where installing more traditional wired or wireless infrastructures is not cost-effective \cite{svitak}.

Since satellite broadband has not yet experienced serious congestion problems, its pricing plans remain fairly simple, though the growth in demand for high-bandwidth services may open up possibilities for new pricing schemes in the future. Technological capabilities that allow for fine granular capacity provisioning and dynamic channel allocation will also facilitate the adoption of more innovative pricing practices. Sun and Modiano \citeyear{sun2006channel} proposed a pricing-based solution to the problem of channel allocation in slotted Aloha-based satellite networks.  
Today, satellite broadband consumers mostly pay in proportion to the uplink and downlink data rates by subscribing to a chosen tier of available data plans. Each tier has specified maximum upload and download speeds and a monthly cap on bandwidth. Beyond this cap, a user's speed is either throttled down or charged by overage, depending on the plan. Table \ref{tab:satellite} gives an overview of key features of different plans, which vary by the granularity of data caps, the offered uplink and downlink speeds, and the type of overage penalties.

\begin{table}[t]
\renewcommand{\arraystretch}{1.3}
\tbl{Satellite plan features and congestion control mechanisms.}{
\centering
\begin{tabular}{ll}
\hline
Feature & Options\\
\hline \hline
\multirow{2}{*}{Data caps (limits)} & Monthly \cite{SA-sat} \\
& Daily \\ 
Uplink and downlink speeds & Varies by company  \cite{viasat,Russian-sat}\\ 
\multirow{2}{*}{Overage penalty} & Throttling to lower speeds \cite{viasat}\\
& Usage-based charges \cite{Russian-sat,avanti} \\
\hline
\end{tabular}}\label{tab:satellite}
\end{table}

\subsection{Pricing of Heterogeneous Networks}

As demand for mobile data grows, ISPs are not only introducing new pricing plans but are also encouraging users to offload their traffic onto other wireless networks. For instance, AT\&T deployed WiFi hotspots in New York and San Francisco after their cellular networks proved unable to handle the demand for data in such congested urban areas \cite{nycwifi}. Also in the United States, Verizon has stated that it expects to run out of LTE spectrum as soon as 2013 in some markets; the company is encouraging consumer adoption of femtocells to supplement this limited capacity \cite{verizonfemto}.  ISPs' key reasons for rolling out small cells are that they help increase capacity supply while keeping costs low and rely on end-user backhauling for data transfers. Moreover, higher band spectrum is now becoming available, which is suitable for small cells, in particular femtocells.  Although small cells are cheap to deploy and work as plug-and-play devices, they can often lead to poor performance due to interference on the shared spectrum, which poses considerable technical challenge. 
With the increasing adoption of small cells and their co-existence with other wireless networks (e.g., 3G, 4G), ISPs will soon need to consider the pricing of heterogeneous networks with devices being able to connect to either of the available network technologies.

As with all pricing, heterogeneous network pricing can be used to steer users to an optimal allocation of resources that maximizes the efficiency of the networks considered. For instance, Neely et al. \citeyear{neely2005fairness} examined flow control in a heterogeneous network setting with wired and wireless services. They employed methods from stochastic optimal control to optimize user utility subject to the bandwidth capacity constraints, an approach that may be interpreted as a type of dynamic market pricing. Chan et al. \citeyear{chan2005utility} assumed that users have access to multiple networks, possibly more than two, and used dynamic pricing to steer users to the ``optimal'' network so as to maximize overall user utility.

Other, more recent works on heterogeneous networks have emphasized the profit and revenue of Internet service providers, rather than employing prices solely to optimize resource management. These works often begin by modeling user adoption of heterogeneous technologies as a function of the access prices; users are assumed to adopt different technologies so as to maximize their utility, which in turn depends on the total number of users adopting each technology (e.g., due to positive externalities of connectivity, or negative externalities from network congestion). Shetty et al. \citeyear{shetty2009economics} used such a model to study revenue maximization for femtocells, while Ren et al. \citeyear{ren2012entry} examined the selection of femtocell spectrum-sharing schemes and the corresponding optimal prices. Joe-Wong et al. \citeyear{wifi-3g-infocom} took a more generic approach, studying the adoption of generic ``base'' and ``supplemental'' technologies and the ISP's profit-maximizing prices when offloading benefits and network deployment costs are considered. A game-theoretic formulation that does not include adoption modeling was proposed by Niyato and Hossain \citeyear{niyato2008competitive}, in which two companies competing for end users were assumed to offer WiFi and WIMAX connectivity.

\subsection{New Pricing Architectures}


Academics have often argued that architectural issues in Internet pricing are more important than the form or basis of the actual prices charged: for instance, Shenker et al. \citeyear{Shenker96} opined that the flat- versus usage-based pricing debate oversimplifies the issue, as in reality there is a continuum between the two.  The authors made the case that research into Internet pricing has overly focused on optimality, and in particular on matching prices to the marginal congestion costs.  Instead, they advocated an architectural focus, i.e., designing the network architecture to facilitate various pricing plans, such as allowing receivers rather than senders to be charged for usage. 
Architecture and systems enabling such pricing innovations for broadband data are arguably becoming an active area of research, and are central to shaping the future research agenda. Understanding these interactions between economics and technology in order to create the overall design of future networks has been identified as a priority in the NSF's Future Internet Design initiative \cite{Darleen-NSF}.    

\begin{figure}
\centering
\includegraphics[width = 0.65\textwidth]{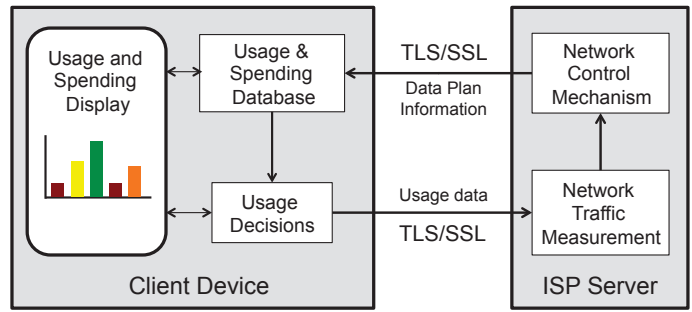}
\caption{Pricing architecture with client- and server-side components.}
\label{fig:arch_tdp}
\end{figure}

As new pricing plans are introduced, ISPs must confront the challenges of ensuring that they are acceptable to users, while also maintaining system scalability, privacy, and security. One way to do this, e.g., as proposed by MacKie-Mason et al. \citeyear{MacKie-Mason}, is to add a client-side component to the current pricing architecture, as shown in Fig. \ref{fig:arch_tdp}. The client-side component includes modules to show users their data consumption and spending over the month, thus allowing users to monitor and control their spending, even if they are unfamiliar with their data plan specifics. Secure connections with the ISP server ensure that user privacy is not compromised. 
In fact, data monitoring apps that sit exclusively on the client side are already being offered on both wireless and wired platforms \cite{datawiz,netlimiter} to help users manage monthly data caps. On wired desktop platforms, users can even control their data usage by controlling the bandwidth rates of different applications. Research on in-home Internet usage has showed that such applications enhance user awareness of data usage and that there is a consumer demand for tools that help them understand their data usage \cite{chetty2010s,chetty2012you}. However, the interfaces of these applications for both wired and wireless networks must be carefully designed in order to facilitate user engagement \cite{chetty2011my,sigchi}.

Client-side usage monitoring applications can go far beyond monitoring usage, especially if integrated with an ISP pricing plan. For instance, the client-side architecture for dynamic pricing can incorporate a module to display current and, if known, future prices as well as usage and spending. In fact, if future prices are known, e.g., day-ahead dynamic pricing, one can even go a step further and incorporate a module that automatically schedules some delay-tolerant applications to less expensive times of the day \cite{ha2012tube}. Such an ``autopilot mode'' for usage decisions incorporates a form of bandwidth control into wireless devices, and integrating it with the prices offered can help make dynamic pricing viable for ISPs.

\begin{table}[t]
\renewcommand{\arraystretch}{1.3}
\tbl{Summary of some emerging trends.}{
\centering
\begin{tabular}{ l l l }
\hline
Trend & Description & Reference\\
\hline
Flat-rate \& usage- &  Coexistence of both types of plans. & \\
based pricing & Emergence of shared data plans. & \cite{shared-Vrz,ATD}\\
& Focus on architectural issues in   & \cite{Shenker96}\\
& Internet pricing is needed. & \\
&&\\
Time-dependent pricing & Traffic shaping with off-peak usage incentives. & \cite{BellLabs}\\
& Static discounts (two-period ToD) &\\
&&\\
Dynamic pricing & Practiced for voice calls (e.g., location-based). & \cite{Economist}\\
& Adopted in Africa \& India. &  \\
& Needs new theory for dynamic data pricing. & \cite{Carlee}\\
& New architecture \& systems implementation. &\cite{Ha}\\
&&\\
Differentiated, &  Tiered data plans with various QoS. & \cite{Ericsson}\\
app-based pricing & App-specific ``zero-rating,'' app bundles. & \cite{tollfreeatt}\\
\hline
\end{tabular}}\label{trends}
\end{table}

\subsection{Emerging Pricing Plans}

A new pricing option that US ISPs are offering is a ``shared data plan,'' which allows users to share their data cap across multiple devices, at a premium for each additional device \cite{shared-Vrz,data-shared}.  Rogers in Canada has been offering such plans on a promotional basis since 2011 \cite{Rogers}. Orange of France Telecom has also started such a plan in Austria that allows iPad owners to share one allotment of data with a phone, and 38\% of iPad owners on its network now subscribe to this plan \cite{ATD}. Sharing a data cap across multiple devices may induce more efficient usage of the quota per consumer. However, an open research question is to understand shared data plans' impact on usage and quota dynamics, with field trials to understand the trade-offs between fairly sharing the caps across devices.
 
In dealing with the growing problem of bandwidth demand, researchers from Alcatel-Lucent have recently proposed static time-dependent discounts for incentivizing off-peak capacity usage \cite{BellLabs}. They also presented policy and QoS mechanisms to implement traffic shaping with off-peak usage incentives to users. Another interesting trend emerging in several parts of the world, particularly in Asia and Africa, is that of a larger scale adoption of dynamic pricing for voice calls \cite{Economist}. As the demand for data, and in particular mobile broadband, grows, there will be a need for new architectures and systems, as reported by Ha et al. \citeyear{Ha} and Joe-Wong et al. \citeyear{Carlee} in the context of pilot-trial results for a dynamic time-dependent usage-based pricing system called TUBE (Time-dependent Usage-based Broadband price Engineering) \cite{TUBEWeb,ha2012tube}. 

Differentiated pricing for broadband \cite{Ericsson}, e.g., application-based pricing of data, is also a trend that is likely to receive more attention from industry. As noted previously, ISPs are exploring new ways to attract customers by ``zero-rating'' or bundling certain applications. A related direction for future research is that of data plans with sponsored content in which user's access fees are subsidized by the content providers or advertisers. By offering such plans, content providers can incentivize users to consume their content, and thus increase their advertising revenue. Sponsored content pricing can be modeled as a two-sided market; using game theory, Andrews et al. \citeyear{andrewsSDP} derived the optimal pricing policies for an ISP. However, such a pricing scheme could be viewed as non-neutral, which may hinder its adoption. To avoid this issue, Sheng et al. \citeyear{workshop} suggested a platform that allows content providers to subsidize a certain dollar amount of users' traffic, which need not correspond to traffic from that content provider. In exchange, users agree to view advertisements, which can help generate content provider revenue. 

\noindent A summary of these potential future trends is given in Table \ref{trends}. 
%
%
%

\section{Conclusions}\label{conclusions}
 
In this work, we draw attention to the growing problem of network congestion and highlight some of the recent steps taken by ISPs to mitigate its effects. The projected growth in demand for data, especially from mobile data and video traffic, is far more than what can be supported even by the latest technological advances, e.g., 4G/LTE and WiFi offloading, due to expensive backhauling and increasing wired network congestion.  Consequently, ISPs have been aggressively using pricing as a congestion control tool.  

The basic idea of congestion pricing has been known in the networking community for several decades, but only now have conditions demanded that it be put into practice. In this survey, we first outline the various threats to the sustainability and economic viability of the Internet ecosystem from the perspectives of ISPs, consumers, and content providers. We then highlight the role that pricing can play in alleviating network congestion and in balancing the functionality goals of data networks, followed by a detailed discussion of the technological, socioeconomic, and regulatory challenges to pricing innovation and related open problems.  We review many known data network pricing proposals, both static and dynamic, and discuss the extent to which some of these have been adopted by ISPs, drawing parallels to existing pricing practices in electricity markets and road networks. We also discuss the predominant trends in access pricing and the need for new architecture, field trials, and interdisciplinary research for realizing more innovative dynamic pricing solutions such as day-ahead time- and usage-dependent pricing, app-based pricing, etc.

The material presented in this paper seeks to inform networking researchers about the existing works on access pricing, the ongoing developments in pricing plans, and the research challenges that need to be accounted for in shaping a new research agenda on smart data pricing.
  
\section*{Acknowledgements}

This work has benefited immensely from our discussions with several colleagues and collaborators.  We would like to specially thank (in no particular order) Victor Glass, Prashanth Hande, Krishan Sabnani, Raj Savoor, Steve Sposato, Rob Calderbank, Danny Tsang, Sundeep Rangan, Yuan Wu, Junshan Zhang, T. Russell Hsing, Keith Cambron, Andrew Odlyzko, Roch Gu\'{e}rin, Augustin Chaintreau and  Jennifer Rexford for their valuable comments, viewpoints, and information that have helped to shape this paper. Additionally, we acknowledge the support of our industrial collaborators, in particular, Reliance Communications of India, MTA of Alaska, SES, AT\&T, Qualcomm, Comcast, and NECA (the National Exchange Carrier Association). We also acknowledge the generous feedback received from the participants of the first Workshop on Smart Data Pricing \cite{SDP}, held in Princeton, NJ on July 30-31, 2012.

\bibliographystyle{acmsmall}
\bibliography{PricingBiblio}

\begin{thebibliography}{}

\bibitem[\protect\citeauthoryear{}{SDP}{2012}]{SDP}
 2012.
\newblock {Smart Data Pricing Forum}.
\newblock http://scenic.princeton.edu/SDP2012.

\bibitem[\protect\citeauthoryear{WPIN}{WPIN}{2012}]{WPIN}
 2012.
\newblock {Workshop on Pricing and Incentives in Networks}.
\newblock http://www.eurecom.fr/~loiseau/W-PIN2012/.

\bibitem[\protect\citeauthoryear{NetEcon}{NetEcon}{2013}]{NetEcon}
 2013.
\newblock {Workshop on the Economics of Networks, Systems and Computation}.
\newblock http://netecon.seas.harvard.edu/.

\bibitem[\protect\citeauthoryear{Al-Manthari, Nasser, and
  Hassanein}{Al-Manthari et~al\mbox{.}}{2011}]{al2011congestion}
{\sc Al-Manthari, B.}, {\sc Nasser, N.}, {\sc and} {\sc Hassanein, H.} 2011.
\newblock Congestion pricing in wireless cellular networks.
\newblock {\em IEEE Communications Surveys \& Tutorials\/}~{\em 13,\/}~3,
  358--371.

\bibitem[\protect\citeauthoryear{{Allot MobileTrends}}{{Allot
  MobileTrends}}{2010}]{allot}
{\sc {Allot MobileTrends}}. 2010.
\newblock {Global Mobile Broadband Traffic Report}.

\bibitem[\protect\citeauthoryear{Anania and Solomon}{Anania and
  Solomon}{1997}]{anania}
{\sc Anania, L.} {\sc and} {\sc Solomon, R.~J.} 1997.
\newblock Flat -- the minimalist price.
\newblock In {\em Internet Economics}, {L.~W. McKnight} {and} {J.~P. Bailey},
  Eds. {The MIT Press}, Cambridge, MA, 91--118.

\bibitem[\protect\citeauthoryear{Andrews, {\"O}zen, Reiman, and Wang}{Andrews
  et~al\mbox{.}}{2013}]{andrewsSDP}
{\sc Andrews, M.}, {\sc {\"O}zen, U.}, {\sc Reiman, M.~I.}, {\sc and} {\sc
  Wang, Q.} 2013.
\newblock Economic models of sponsored content in wireless networks with
  uncertain demand.
\newblock In {\em Proceedings of the Second Smart Data Pricing Workshop}. IEEE.

\bibitem[\protect\citeauthoryear{Apple}{Apple}{2012}]{apple}
{\sc Apple}. 2012.
\newblock Apple--the new {iPad}.
\newblock http://www.apple.com/ipad/compare/.

\bibitem[\protect\citeauthoryear{Borenstein, Jaske, and Rosenfeld}{Borenstein
  et~al\mbox{.}}{2002}]{borenstein2002dynamic}
{\sc Borenstein, S.}, {\sc Jaske, M.}, {\sc and} {\sc Rosenfeld, A.} 2002.
\newblock Dynamic pricing, advanced metering, and demand response in
  electricity markets.
\newblock Tech. rep., Center for the Study of Energy Markets.
\newblock Working Paper.

\bibitem[\protect\citeauthoryear{Breker}{Breker}{1996}]{Breker}
{\sc Breker, L.~P.} 1996.
\newblock A survey of network pricing schemes.
\newblock In {\em Proceedings of the 8th Symposium on Computer Science}.
  University of Saskatchewan.

\bibitem[\protect\citeauthoryear{Brownlee}{Brownlee}{1997}]{nevil}
{\sc Brownlee, N.} 1997.
\newblock Internet pricing in practice.
\newblock In {\em Internet Economics}, {L.~W. McKnight} {and} {J.~P. Bailey},
  Eds. {The MIT Press}, Cambridge, MA, 77--90.

\bibitem[\protect\citeauthoryear{Bubley}{Bubley}{2011}]{Mobistar}
{\sc Bubley, D.} 2011.
\newblock Belgian {MNO} tries app-specific zero rating - bad idea in my view.
\newblock Disruptive Wireless.
\newblock November 1, http://tinyurl.com/6urunl3.

\bibitem[\protect\citeauthoryear{Buckley}{Buckley}{2012}]{shared-Vrz}
{\sc Buckley, S.} 2012.
\newblock {Verizon shared data plans show up in employee training materials,
  still on track?}
\newblock Engadget.
\newblock January 30.

\bibitem[\protect\citeauthoryear{Carew}{Carew}{2012}]{att-tsunami}
{\sc Carew, S.} 2012.
\newblock {Users complain, AT\&T blames data tsunami}.
\newblock Reuters Media File.
\newblock February 14.

\bibitem[\protect\citeauthoryear{Caron and Kesidis}{Caron and
  Kesidis}{2010}]{caron2010incentive}
{\sc Caron, S.} {\sc and} {\sc Kesidis, G.} 2010.
\newblock Incentive-based energy consumption scheduling algorithms for the
  smart grid.
\newblock In {\em Proceedings of the First IEEE International Conference on
  Smart Grid Communications}. IEEE, 391--396.

\bibitem[\protect\citeauthoryear{Chan, Fan, and Cao}{Chan
  et~al\mbox{.}}{2005}]{chan2005utility}
{\sc Chan, H.}, {\sc Fan, P.}, {\sc and} {\sc Cao, Z.} 2005.
\newblock A utility-based network selection scheme for multiple services in
  heterogeneous networks.
\newblock In {\em Proceedings of the International Conferences on Wireless
  Networks, Communications and Mobile Computing}. Vol.~2. IEEE, 1175--1180.

\bibitem[\protect\citeauthoryear{Chang and Petr}{Chang and
  Petr}{2001}]{chang2001survey}
{\sc Chang, X.} {\sc and} {\sc Petr, D.} 2001.
\newblock A survey of pricing for integrated service networks.
\newblock {\em Computer Communications\/}~{\em 24,\/}~18, 1808--1818.

\bibitem[\protect\citeauthoryear{{Charles River Associates}}{{Charles River
  Associates}}{2005}]{CharlesRiver}
{\sc {Charles River Associates}}. 2005.
\newblock Impact evaluation of the {C}alifornia statewide pricing pilot.
\newblock Tech. rep., Charles River Associates.

\bibitem[\protect\citeauthoryear{Chen}{Chen}{2012}]{nyt-shared}
{\sc Chen, B.~X.} 2012.
\newblock {Shared mobile data plans: Who benefits?}
\newblock New York Times.
\newblock July 19, Bits Blog.

\bibitem[\protect\citeauthoryear{Chen, Ghosh, Magutt, and Chiang}{Chen
  et~al\mbox{.}}{2012}]{Jiasi}
{\sc Chen, J.}, {\sc Ghosh, A.}, {\sc Magutt, J.}, {\sc and} {\sc Chiang, M.}
  2012.
\newblock {QAVA}: Quota aware video adaptation.
\newblock In {\em Proceedings of ACM CoNEXT}. ACM, 121--132.

\bibitem[\protect\citeauthoryear{Chen and Jana}{Chen and Jana}{2013}]{chen2013}
{\sc Chen, Y.-F.~R.} {\sc and} {\sc Jana, R.} 2013.
\newblock {SpeedGate: A} smart data pricing testbed based on speed tiers.
\newblock In {\em Proceedings of the Second Smart Data Pricing Workshop}. IEEE.

\bibitem[\protect\citeauthoryear{Chetty, Banks, Brush, Donner, and
  Grinter}{Chetty et~al\mbox{.}}{2012}]{chetty2012you}
{\sc Chetty, M.}, {\sc Banks, R.}, {\sc Brush, A.}, {\sc Donner, J.}, {\sc and}
  {\sc Grinter, R.} 2012.
\newblock You're capped: {U}nderstanding the effects of bandwidth caps on
  broadband use in the home.
\newblock In {\em Proceedings of ACM CHI}. ACM, 3021--3030.

\bibitem[\protect\citeauthoryear{Chetty, Banks, Harper, Regan, Sellen,
  Gkantsidis, Karagiannis, and Key}{Chetty et~al\mbox{.}}{2010}]{chetty2010s}
{\sc Chetty, M.}, {\sc Banks, R.}, {\sc Harper, R.}, {\sc Regan, T.}, {\sc
  Sellen, A.}, {\sc Gkantsidis, C.}, {\sc Karagiannis, T.}, {\sc and} {\sc Key,
  P.} 2010.
\newblock Who's hogging the bandwidth: {T}he consequences of revealing the
  invisible in the home.
\newblock In {\em Proceedings of ACM CHI}. ACM, 659--668.

\bibitem[\protect\citeauthoryear{Chetty, Haslem, Baird, Ofoha, Sumner, and
  Grinter}{Chetty et~al\mbox{.}}{2011}]{chetty2011my}
{\sc Chetty, M.}, {\sc Haslem, D.}, {\sc Baird, A.}, {\sc Ofoha, U.}, {\sc
  Sumner, B.}, {\sc and} {\sc Grinter, R.} 2011.
\newblock Why is my {I}nternet slow?: {M}aking network speeds visible.
\newblock In {\em Proceedings of ACM CHI}. ACM, 1889--1898.

\bibitem[\protect\citeauthoryear{{Cisco Systems}}{{Cisco
  Systems}}{2012a}]{CiscoVNI}
{\sc {Cisco Systems}}. 2012a.
\newblock Cisco visual networking index: {F}orecast and methodology, 2011-2016.
\newblock May 30, http://tinyurl.com/VNI2011.

\bibitem[\protect\citeauthoryear{{Cisco Systems}}{{Cisco
  Systems}}{2012b}]{CiscoVNImobile}
{\sc {Cisco Systems}}. 2012b.
\newblock Cisco visual networking index: {G}lobal mobile data traffic forecast
  update, 2011-2016.
\newblock February 14, http://tinyurl.com/VNI2012-mobile.

\bibitem[\protect\citeauthoryear{Clark}{Clark}{1997}]{DDClark}
{\sc Clark, D.~D.} 1997.
\newblock Internet cost allocation and pricing.
\newblock In {\em Internet Economics}, {L.~W. McKnight} {and} {J.~P. Bailey},
  Eds. {The MIT Press}, Cambridge, MA, 215--252.

\bibitem[\protect\citeauthoryear{Cocchi, Estrin, Shenker, and Zhang}{Cocchi
  et~al\mbox{.}}{1991}]{Cocchi1}
{\sc Cocchi, R.}, {\sc Estrin, D.}, {\sc Shenker, S.}, {\sc and} {\sc Zhang,
  L.} 1991.
\newblock A study of priority pricing in multiple service class networks.
\newblock {\em ACM SIGCOMM Computer Communication Review\/}~{\em 21,\/}~4,
  123--130.

\bibitem[\protect\citeauthoryear{Cocchi, Shenker, Estrin, and Zhang}{Cocchi
  et~al\mbox{.}}{1993}]{Cocchi2}
{\sc Cocchi, R.}, {\sc Shenker, S.}, {\sc Estrin, D.}, {\sc and} {\sc Zhang,
  L.} 1993.
\newblock Pricing in computer networks: {M}otivation, formulation, and example.
\newblock {\em {IEEE/ACM Transactions on Networking}\/}~{\em 1}, 614--627.

\bibitem[\protect\citeauthoryear{Comcast}{Comcast}{2012}]{Comcast-cap}
{\sc Comcast}. 2012.
\newblock About excessive use of data.
\newblock September,
  http://customer.comcast.com/help-and-support/internet/data-usage-what-are-the-different-plans-launching.

\bibitem[\protect\citeauthoryear{Courcoubetis, Stamoulis, Manolakis, and
  Kelly}{Courcoubetis et~al\mbox{.}}{1998}]{Courcoubetis}
{\sc Courcoubetis, C.}, {\sc Stamoulis, G.~D.}, {\sc Manolakis, C.}, {\sc and}
  {\sc Kelly, F.~P.} 1998.
\newblock An intelligent agent for optimizing {QoS}-for-money in priced {ABR}
  connections.
\newblock Tech. rep., Institute of Computer Science--Foundation for Research
  and Technology Hellas and Statistical Laboratory, University of Cambridge.
\newblock Preprint.

\bibitem[\protect\citeauthoryear{Dash, Kantere, and Ailamaki}{Dash
  et~al\mbox{.}}{2009}]{dash2009economic}
{\sc Dash, D.}, {\sc Kantere, V.}, {\sc and} {\sc Ailamaki, A.} 2009.
\newblock An economic model for self-tuned cloud caching.
\newblock In {\em Proceedings of IEEE ICDE}. IEEE, 1687--1693.

\bibitem[\protect\citeauthoryear{DataMi}{DataMi}{2012}]{TUBEWeb}
{\sc DataMi}. 2012.
\newblock {DataMi Website}.
\newblock {http://scenic.princeton.edu/datami/}.

\bibitem[\protect\citeauthoryear{Delgrossi and Ferrari}{Delgrossi and
  Ferrari}{1999}]{Delgrossi}
{\sc Delgrossi, L.} {\sc and} {\sc Ferrari, D.} 1999.
\newblock Charging schemes for reservation-based networks.
\newblock {\em Telecommunication Systems\/}~{\em 11,\/}~1, 127--137.

\bibitem[\protect\citeauthoryear{D'Orazio}{D'Orazio}{2012}]{viasat}
{\sc D'Orazio, D.} 2012.
\newblock {Viasat prices satellite broadband plans, calls them Exede}.
\newblock TheVerge.
\newblock January 10,
  http://www.theverge.com/2012/1/10/2694708/viasat-exede-pricing-availability-satellite-broadband-internet.

\bibitem[\protect\citeauthoryear{Du and Lu}{Du and Lu}{2011}]{Du}
{\sc Du, P.} {\sc and} {\sc Lu, N.} 2011.
\newblock Appliance commitment for household load scheduling.
\newblock {\em IEEE Transactions on Smart Grid\/}~{\em 2,\/}~2, 411--419.

\bibitem[\protect\citeauthoryear{Dyaberi, Parsons, Pai, Kannan, Chen, Jana,
  Stern, and Varshavsky}{Dyaberi et~al\mbox{.}}{2012}]{dyaberi2012managing}
{\sc Dyaberi, J.~M.}, {\sc Parsons, B.}, {\sc Pai, V.~S.}, {\sc Kannan, K.},
  {\sc Chen, Y.-F.~R.}, {\sc Jana, R.}, {\sc Stern, D.}, {\sc and} {\sc
  Varshavsky, A.} 2012.
\newblock Managing cellular congestion using incentives.
\newblock {\em IEEE Communications Magazine\/}~{\em 50,\/}~11, 100--107.

\bibitem[\protect\citeauthoryear{El-Sayed, Mukhopadhyay, Urrutia-Vald\'{e}s,
  and Zhao}{El-Sayed et~al\mbox{.}}{2011}]{BellLabs}
{\sc El-Sayed, M.}, {\sc Mukhopadhyay, A.}, {\sc Urrutia-Vald\'{e}s, C.}, {\sc
  and} {\sc Zhao, Z.~J.} 2011.
\newblock Mobile data explosion: {M}onetizing the opportunity through dynamic
  policies and {QoS} pipes.
\newblock {\em Bell Labs Technical Journal\/}~{\em 16,\/}~2, 79--100.

\bibitem[\protect\citeauthoryear{Electronista}{Electronista}{2011}]{Rogers}
{\sc Electronista}. 2011.
\newblock {Rogers intros data sharing for 6GB plans with hefty premium}.
\newblock February 8,
  http://www.electronista.com/articles/11/02/08/rogers.6gb.data.plan.gets.expensive.sharing/.

\bibitem[\protect\citeauthoryear{{Electronista}}{{Electronista}}{2012}]{verizonfemto}
{\sc {Electronista}}. 2012.
\newblock {Verizon preps LTE femtocells, says it hits limits in 2013}.
\newblock Electronista.
\newblock http://www.electronista.com/articles/
  12/03/05/demand.for.mobile.data.will.outpace.build.out/.

\bibitem[\protect\citeauthoryear{Ericsson}{Ericsson}{2010}]{japan-ericsson}
{\sc Ericsson}. 2010.
\newblock {Japan's Softbank Mobile chooses Ericsson mobile broadband solution}.
\newblock August 2, Press Release.

\bibitem[\protect\citeauthoryear{Ericsson}{Ericsson}{2011}]{Ericsson}
{\sc Ericsson}. 2011.
\newblock Differentiated mobile broadband.
\newblock White Paper, http://www.ericsson.com/res/docs/whitepapers/
  differentiated\_mobile\_broadband.pdf.

\bibitem[\protect\citeauthoryear{Ezziane}{Ezziane}{2005}]{Ezziane}
{\sc Ezziane, Z.} 2005.
\newblock Charging and pricing challenges for 3{G} systems.
\newblock {\em IEEE Communications Surveys \& Tutorials\/}~{\em 7,\/}~4,
  58--68.

\bibitem[\protect\citeauthoryear{Falkner, Devetsikiotis, and
  Lambadaris}{Falkner et~al\mbox{.}}{2000}]{Falkner}
{\sc Falkner, M.}, {\sc Devetsikiotis, M.}, {\sc and} {\sc Lambadaris, I.}
  2000.
\newblock An overview of pricing concepts for broadband {IP} networks.
\newblock {\em {IEEE} Communications Surveys\/}~{\em 3}, 2--13.

\bibitem[\protect\citeauthoryear{Faruqui}{Faruqui}{2010}]{faruqui2010}
{\sc Faruqui, A.} 2010.
\newblock The ethics of dynamic pricing.
\newblock {\em The Electricity Journal\/}~{\em 23,\/}~6, 13--28.

\bibitem[\protect\citeauthoryear{Faruqui, Hledik, and Sergici}{Faruqui
  et~al\mbox{.}}{2009}]{faruqui2009piloting}
{\sc Faruqui, A.}, {\sc Hledik, R.}, {\sc and} {\sc Sergici, S.} 2009.
\newblock Piloting the smart grid.
\newblock {\em The Electricity Journal\/}~{\em 22,\/}~7, 55--69.

\bibitem[\protect\citeauthoryear{Faruqui and Wood}{Faruqui and
  Wood}{2008}]{faruqui2008quantifying}
{\sc Faruqui, A.} {\sc and} {\sc Wood, L.} 2008.
\newblock Quantifying the benefits of dynamic pricing in the mass market.
\newblock Tech. rep., Edison Electric Institute.

\bibitem[\protect\citeauthoryear{Fisher}{Fisher}{2007}]{Darleen-NSF}
{\sc Fisher, D.} 2007.
\newblock {US National Science Foundation and the Future Internet Design}.
\newblock {\em {ACM SIGCOMM Computer Communication Review}\/}~{\em 37,\/}~3,
  85--87.

\bibitem[\protect\citeauthoryear{Fitchard}{Fitchard}{2012}]{Fitchard}
{\sc Fitchard, K.} 2012.
\newblock New {N}etflix i{OS} app capitulates to bandwidth caps.
\newblock GigaOm.
\newblock May 31,
  http://gigaom.com/mobile/new-netflix-ios-app-capitulates-to-bandwidth-caps/.

\bibitem[\protect\citeauthoryear{Frakes}{Frakes}{2010}]{ATT-usagebased}
{\sc Frakes, D.} 2010.
\newblock {AT\&T announces tethering details and new plans for iPhone, iPad}.
\newblock InfoWorld.

\bibitem[\protect\citeauthoryear{Fried}{Fried}{2011}]{ATD}
{\sc Fried, I.} 2011.
\newblock The data plan of the future is available now, at least in {E}urope.
\newblock All Things D.
\newblock http://tinyurl.com/FT-shared.

\bibitem[\protect\citeauthoryear{Galbraith}{Galbraith}{2011}]{data-shared}
{\sc Galbraith, C.} 2011.
\newblock {Shared data plans set to make global impact}.
\newblock Billing World.
\newblock October 24.

\bibitem[\protect\citeauthoryear{Ganesh, Laevens, and Steinberg}{Ganesh
  et~al\mbox{.}}{2001}]{Ganesh}
{\sc Ganesh, A.}, {\sc Laevens, K.}, {\sc and} {\sc Steinberg, R.} 2001.
\newblock Congestion pricing and user adaptation.
\newblock In {\em Proceedings of IEEE INFOCOM}. Vol.~2. IEEE, 959--965.

\bibitem[\protect\citeauthoryear{Genachowski}{Genachowski}{2010}]{FCC-JG}
{\sc Genachowski, J.} 2010.
\newblock New rules for an open {I}nternet.
\newblock US Federal Communications Commission.
\newblock December 21, http://www.fcc.gov/blog/new-rules-open-internet.

\bibitem[\protect\citeauthoryear{Gizelis and Vergados}{Gizelis and
  Vergados}{2011}]{Gizelis}
{\sc Gizelis, C.} {\sc and} {\sc Vergados, D.} 2011.
\newblock A survey of pricing schemes in wireless networks.
\newblock {\em IEEE Communications Surveys \& Tutorials\/}~{\em 13,\/}~1,
  126--145.

\bibitem[\protect\citeauthoryear{Glass, Prinzivalli, and Stefanova}{Glass
  et~al\mbox{.}}{2012}]{vglass-2}
{\sc Glass, V.}, {\sc Prinzivalli, J.}, {\sc and} {\sc Stefanova, S.} 2012.
\newblock Persistence of middle mile problems for rural exchanges local
  carriers.
\newblock Smart Data Pricing Workshop Talk.
\newblock http://scenic.princeton.edu/SDP2012/Talks-VictorGlass.pdf.

\bibitem[\protect\citeauthoryear{Goldstein}{Goldstein}{2012}]{tollfreeatt}
{\sc Goldstein, P.} 2012.
\newblock {AT\&T's S}tephenson: {C}ontent providers are asking for 'toll free'
  data plans.
\newblock Fierce{W}ireless.
\newblock
  http://www.fiercewireless.com/story/atts-stephenson-content-providers-are-asking-toll-free-data-plans/2012-06-01\#ixzz2M466tpzz.

\bibitem[\protect\citeauthoryear{Gomez-Ibanez and Small}{Gomez-Ibanez and
  Small}{1994}]{RoadPricing}
{\sc Gomez-Ibanez, J.~A.} {\sc and} {\sc Small, K.~A.} 1994.
\newblock {\em Road pricing for congestion management: {A} survey of
  international practice}.
\newblock Transportation Research Board, Washington, DC.
\newblock National Cooperative Highway Research Program.

\bibitem[\protect\citeauthoryear{Gupta, Stahl, and Whinston}{Gupta
  et~al\mbox{.}}{1997}]{Gupta}
{\sc Gupta, A.}, {\sc Stahl, D.}, {\sc and} {\sc Whinston, A.} 1997.
\newblock Priority pricing of integrated services networks.
\newblock In {\em Internet Economics}, {L.~W. McKnight} {and} {J.~P. Bailey},
  Eds. {The MIT Press}, Cambridge, MA, 323--352.

\bibitem[\protect\citeauthoryear{Ha, Joe-Wong, Sen, and Chiang}{Ha
  et~al\mbox{.}}{2012a}]{Ha}
{\sc Ha, S.}, {\sc Joe-Wong, C.}, {\sc Sen, S.}, {\sc and} {\sc Chiang, M.}
  2012a.
\newblock Pricing by timing: {I}nnovating broadband data plans.
\newblock In {\em Proceedings of the SPIE OPTO Broadband Access Communication
  Technologies VI Conference}.
\newblock Paper 8282-12.

\bibitem[\protect\citeauthoryear{Ha, Sen, Joe-Wong, Im, and Chiang}{Ha
  et~al\mbox{.}}{2012b}]{ha2012tube}
{\sc Ha, S.}, {\sc Sen, S.}, {\sc Joe-Wong, C.}, {\sc Im, Y.}, {\sc and} {\sc
  Chiang, M.} 2012b.
\newblock {TUBE: T}ime-dependent pricing for mobile data.
\newblock In {\em Proceedings of ACM SIGCOMM}. ACM, 247--258.

\bibitem[\protect\citeauthoryear{Hande, Chiang, Calderbank, and Zhang}{Hande
  et~al\mbox{.}}{2010}]{hande}
{\sc Hande, P.}, {\sc Chiang, M.}, {\sc Calderbank, R.}, {\sc and} {\sc Zhang,
  J.} 2010.
\newblock Pricing under constraints in access networks: {R}evenue maximization
  and congestion management.
\newblock In {\em Proceedings of IEEE INFOCOM}. IEEE, 1--9.

\bibitem[\protect\citeauthoryear{Harsman}{Harsman}{2001}]{Harsman}
{\sc Harsman, B.} 2001.
\newblock {Urban road pricing acceptance}.
\newblock In {\em IMPRINT-EUROPE Seminar}.

\bibitem[\protect\citeauthoryear{Hausmann, Kinnucan, and McFaddden}{Hausmann
  et~al\mbox{.}}{1979}]{hausmann1979two}
{\sc Hausmann, J.}, {\sc Kinnucan, M.}, {\sc and} {\sc McFaddden, D.} 1979.
\newblock A two-level electricity demand model: {E}valuation of the
  {C}onnecticut time-of-day pricing test.
\newblock {\em Journal of Econometrics\/}~{\em 10,\/}~3, 263--289.

\bibitem[\protect\citeauthoryear{Hayel and Tuffin}{Hayel and
  Tuffin}{2005}]{Hayel}
{\sc Hayel, Y.} {\sc and} {\sc Tuffin, B.} 2005.
\newblock A mathematical analysis of the cumulus pricing scheme.
\newblock {\em {Computer Networks}\/}~{\em 47}, 907--921.

\bibitem[\protect\citeauthoryear{Hayer}{Hayer}{1993}]{Hayer}
{\sc Hayer, J.} 1993.
\newblock Transportation auction: {A} new service concept.
\newblock M.S.\ thesis, University of Alberta.
\newblock {TR-93-05}.

\bibitem[\protect\citeauthoryear{Hendershott}{Hendershott}{2006}]{Hendershott}
{\sc Hendershott, T.} 2006.
\newblock {\em Economics and Information Systems, Volume 1}.
\newblock Elsevier, Amsterdam, Chapter~2, 4--9.

\bibitem[\protect\citeauthoryear{Herter}{Herter}{2007}]{herter2007residential}
{\sc Herter, K.} 2007.
\newblock Residential implementation of critical-peak pricing of electricity.
\newblock {\em Energy Policy\/}~{\em 35,\/}~4, 2121--2130.

\bibitem[\protect\citeauthoryear{Higginbotham}{Higginbotham}{2010}]{Gigaom}
{\sc Higginbotham, S.} 2010.
\newblock Mobile operators want to charge based on time and apps.
\newblock GigaOm.
\newblock December 14.

\bibitem[\protect\citeauthoryear{Holguin-Veras, Cetin, and Xia}{Holguin-Veras
  et~al\mbox{.}}{2006}]{holguin2006comparative}
{\sc Holguin-Veras, J.}, {\sc Cetin, M.}, {\sc and} {\sc Xia, S.} 2006.
\newblock A comparative analysis of us toll policy.
\newblock {\em Transportation Research Part A: Policy and Practice\/}~{\em
  40,\/}~10, 852--871.

\bibitem[\protect\citeauthoryear{Hosanagar, Krishnan, Chuang, and
  Choudhary}{Hosanagar et~al\mbox{.}}{2005}]{hosanagar2005pricing}
{\sc Hosanagar, K.}, {\sc Krishnan, R.}, {\sc Chuang, J.}, {\sc and} {\sc
  Choudhary, V.} 2005.
\newblock Pricing and resource allocation in caching services with multiple
  levels of quality of service.
\newblock {\em Management Science\/}~{\em 51,\/}~12, 1844--1859.

\bibitem[\protect\citeauthoryear{{I.B.M.}}{{I.B.M.}}{2007}]{board2007ontario}
{\sc {I.B.M.}} 2007.
\newblock Ontario energy board smart price pilot final report.
\newblock Tech. rep., Ontario Energy Board.
\newblock {I. B. M. Global Business Services and eMeter Strategic Consulting}.

\bibitem[\protect\citeauthoryear{{IPNet JSC}}{{IPNet JSC}}{2012}]{Russian-sat}
{\sc {IPNet JSC}}. 2012.
\newblock Price and rates.
\newblock http://www.zaoipnet.ru/en/price/connectionservice/.

\bibitem[\protect\citeauthoryear{Jiang, Parekh, and Walrand}{Jiang
  et~al\mbox{.}}{2008}]{Jiang}
{\sc Jiang, L.}, {\sc Parekh, S.}, {\sc and} {\sc Walrand, J.} 2008.
\newblock Time-dependent network pricing and bandwidth trading.
\newblock In {\em Proceedings of the IEEE Network Operations and Management
  Symposium Workshop}. IEEE, 193--200.

\bibitem[\protect\citeauthoryear{Joe-Wong, Ha, and Chiang}{Joe-Wong
  et~al\mbox{.}}{2011}]{Carlee}
{\sc Joe-Wong, C.}, {\sc Ha, S.}, {\sc and} {\sc Chiang, M.} 2011.
\newblock Time-dependent broadband pricing: {F}easibility and benefits.
\newblock In {\em Proceedings of the 31st IEEE ICDCS}. IEEE, 288--298.

\bibitem[\protect\citeauthoryear{Joe-Wong, Sen, and Ha}{Joe-Wong
  et~al\mbox{.}}{2013}]{wifi-3g-infocom}
{\sc Joe-Wong, C.}, {\sc Sen, S.}, {\sc and} {\sc Ha, S.} 2013.
\newblock Offering supplementary wireless network technologies: Adoption
  behavior and offloading benefits.
\newblock In {\em Proceedings of IEEE INFOCOM}.
\newblock http://arxiv.org/abs/1209.5004.

\bibitem[\protect\citeauthoryear{Joe-Wong, Sen, Ha, and Chiang}{Joe-Wong
  et~al\mbox{.}}{2012a}]{joe2012optimized}
{\sc Joe-Wong, C.}, {\sc Sen, S.}, {\sc Ha, S.}, {\sc and} {\sc Chiang, M.}
  2012a.
\newblock Optimized day-ahead pricing for smart grids with device-specific
  scheduling flexibility.
\newblock {\em IEEE Journal on Selected Areas in Communications\/}~{\em
  30,\/}~6, 1075--1085.

\bibitem[\protect\citeauthoryear{Joe-Wong, Sen, Lan, and Chiang}{Joe-Wong
  et~al\mbox{.}}{2012b}]{joe2012multi}
{\sc Joe-Wong, C.}, {\sc Sen, S.}, {\sc Lan, T.}, {\sc and} {\sc Chiang, M.}
  2012b.
\newblock Multi-resource allocation: Fairness-efficiency tradeoffs in a
  unifying framework.
\newblock In {\em Proceedings of IEEE INFOCOM}. IEEE, 1206--1214.

\bibitem[\protect\citeauthoryear{Kang}{Kang}{2010}]{ATT-mobile}
{\sc Kang, C.} 2010.
\newblock {{AT\&T} wireless scraps flat-rate {I}nternet plan}.
\newblock The Washington Post.
\newblock June 2.

\bibitem[\protect\citeauthoryear{Keall}{Keall}{2011}]{TelstraClear}
{\sc Keall, C.} 2011.
\newblock Telstra{C}lear calls limitless weekend `very successful',
  {InternetNZ} `disastrous'.
\newblock The National Business Review.
\newblock December 5.

\bibitem[\protect\citeauthoryear{Kelly}{Kelly}{1994}]{Kelly-eff}
{\sc Kelly, F.} 1994.
\newblock On tariffs, policing, and admissions control for multiservice
  networks.
\newblock {\em Operations Research Letters\/}~{\em 15}, 1--9.

\bibitem[\protect\citeauthoryear{Kelly, Maulloo, and Tan}{Kelly
  et~al\mbox{.}}{1998}]{Kelly-PFP}
{\sc Kelly, F.}, {\sc Maulloo, A.~K.}, {\sc and} {\sc Tan, D. H.~K.} 1998.
\newblock Rate control for communication networks: {S}hadow prices,
  proportional fairness, and stability.
\newblock {\em Journal of the Operational Research Society\/}~{\em 49},
  237--252.

\bibitem[\protect\citeauthoryear{Key}{Key}{2010}]{Comcast-Level3}
{\sc Key, P.} 2010.
\newblock Comcast, {L}evel 3, {N}etflix, the {FCC}: Busy week for neutrality
  debate.
\newblock Philadelphia Business Journal.
\newblock December 1.

\bibitem[\protect\citeauthoryear{Klopfenstein}{Klopfenstein}{2009}]{Klopfenstein}
{\sc Klopfenstein, B.~C.} 2009.
\newblock Internet economics: An annotated bibliography.
\newblock {\em Journal of Media Economics\/}~{\em 11,\/}~1, 33--48.

\bibitem[\protect\citeauthoryear{Kwang}{Kwang}{2011}]{SingTel}
{\sc Kwang, K.} 2011.
\newblock Sing{T}el unveils priority mobile broadband access.
\newblock June 13, http://www.zdnetasia.com/
  singtel-unveils-priority-mobile-broadband-access-62300719.htm.

\bibitem[\protect\citeauthoryear{Lam, Chiu, and Lui}{Lam
  et~al\mbox{.}}{2007}]{Lam}
{\sc Lam, K.~R.}, {\sc Chiu, D.~M.}, {\sc and} {\sc Lui, J. C.~S.} 2007.
\newblock On the access pricing and network scaling issues of wireless mesh
  networks.
\newblock {\em {IEEE} Transactions on Computing\/}~{\em 56}, 140--146.

\bibitem[\protect\citeauthoryear{Laoutaris, Sirivianos, Yang, and
  Rodriguez}{Laoutaris et~al\mbox{.}}{2011}]{laoutaris2011inter}
{\sc Laoutaris, N.}, {\sc Sirivianos, M.}, {\sc Yang, X.}, {\sc and} {\sc
  Rodriguez, P.} 2011.
\newblock Inter-datacenter bulk transfers with netstitcher.
\newblock {\em ACM SIGCOMM Computer Communication Review\/}~{\em 41,\/}~4,
  74--85.

\bibitem[\protect\citeauthoryear{Lee, Mo, Walrand, and Park}{Lee
  et~al\mbox{.}}{2011}]{Token}
{\sc Lee, D.}, {\sc Mo, J.}, {\sc Walrand, J.}, {\sc and} {\sc Park, J.} 2011.
\newblock A token pricing scheme for {I}nternet services.
\newblock In {\em Proceedings of the Seventh ICQT}.

\bibitem[\protect\citeauthoryear{Lewis}{Lewis}{1996}]{Lewis}
{\sc Lewis, P.~H.} 1996.
\newblock {An ``all you can eat" price is clogging Internet access}.
\newblock New York Times.
\newblock December 17, p. A1.

\bibitem[\protect\citeauthoryear{Li, Chen, and Low}{Li
  et~al\mbox{.}}{2011}]{li2011optimal}
{\sc Li, N.}, {\sc Chen, L.}, {\sc and} {\sc Low, S.} 2011.
\newblock Optimal demand response based on utility maximization in power
  networks.
\newblock In {\em IEEE Power and Energy Society General Meeting}. IEEE, 1--8.

\bibitem[\protect\citeauthoryear{Li, Huang, and Li}{Li
  et~al\mbox{.}}{2009}]{Li}
{\sc Li, S.}, {\sc Huang, J.}, {\sc and} {\sc Li, S. Y.~R.} 2009.
\newblock Revenue maximization for communication networks with usage-based
  pricing.
\newblock In {\em Proceedings of IEEE GLOBECOM}. IEEE, 1--6.

\bibitem[\protect\citeauthoryear{{Locktime Software}}{{Locktime
  Software}}{2012}]{netlimiter}
{\sc {Locktime Software}}. 2012.
\newblock Netlimiter website.
\newblock http://www.netlimiter.com/.

\bibitem[\protect\citeauthoryear{Loiseau, Schwartz, Musacchio, and
  Amin}{Loiseau et~al\mbox{.}}{2011}]{loiseau2011incentive}
{\sc Loiseau, P.}, {\sc Schwartz, G.}, {\sc Musacchio, J.}, {\sc and} {\sc
  Amin, S.} 2011.
\newblock Incentive schemes for {I}nternet congestion management: {R}affles
  versus time-of-day pricing.
\newblock In {\em Proceedings of the Allerton Conference}. IEEE, 103--110.

\bibitem[\protect\citeauthoryear{MacKie-Mason and Varian}{MacKie-Mason and
  Varian}{1995}]{Varian}
{\sc MacKie-Mason, J.} {\sc and} {\sc Varian, H.} 1995.
\newblock Pricing the {I}nternet.
\newblock In {\em Public Access to the Internet}, {B.~Kahin} {and} {J.~Keller},
  Eds. Prentice-Hall, Englewood Cliffs, NJ.

\bibitem[\protect\citeauthoryear{MacKie-Mason, Murphy, and Murphy}{MacKie-Mason
  et~al\mbox{.}}{1997}]{MacKie-Mason}
{\sc MacKie-Mason, J.~K.}, {\sc Murphy, L.}, {\sc and} {\sc Murphy, J.} 1997.
\newblock Responsive pricing in the {I}nternet.
\newblock In {\em Internet Economics}, {L.~W. McKnight} {and} {J.~P. Bailey},
  Eds. {The MIT Press}, Cambridge, MA, 279--303.

\bibitem[\protect\citeauthoryear{Marbach}{Marbach}{2004}]{Marbach}
{\sc Marbach, P.} 2004.
\newblock Analysis of a static pricing scheme for priority services.
\newblock {\em {IEEE/ACM} Transactions on Networking\/}~{\em 12}, 312--325.

\bibitem[\protect\citeauthoryear{Marcon, Santos, Gummadi, Laoutaris, Rodriguez,
  and Vahdat}{Marcon et~al\mbox{.}}{2012}]{marcon2012netex}
{\sc Marcon, M.}, {\sc Santos, N.}, {\sc Gummadi, K.~P.}, {\sc Laoutaris, N.},
  {\sc Rodriguez, P.}, {\sc and} {\sc Vahdat, A.} 2012.
\newblock {NetEx: C}ost-effective bulk data transfers for cloud computing.
\newblock Tech. rep., Max Planck Institute for Software Systems.

\bibitem[\protect\citeauthoryear{Marshall}{Marshall}{1996}]{Marshall}
{\sc Marshall, J.} 1996.
\newblock {Economics, not engineering, will unclog the Internet}.
\newblock San Francisco Chronicle.
\newblock November 4.

\bibitem[\protect\citeauthoryear{Matsukawa}{Matsukawa}{2001}]{matsukawa2001household}
{\sc Matsukawa, I.} 2001.
\newblock Household response to optional peak-load pricing of electricity.
\newblock {\em Journal of Regulatory Economics\/}~{\em 20,\/}~3, 249--267.

\bibitem[\protect\citeauthoryear{May}{May}{1992}]{may1992road}
{\sc May, A.} 1992.
\newblock Road pricing: {A}n international perspective.
\newblock {\em Transportation\/}~{\em 19,\/}~4, 313--333.

\bibitem[\protect\citeauthoryear{May and Milne}{May and
  Milne}{2000}]{may2000effects}
{\sc May, A.} {\sc and} {\sc Milne, D.} 2000.
\newblock Effects of alternative road pricing systems on network performance.
\newblock {\em Transportation Research Part A: Policy and Practice\/}~{\em
  34,\/}~6, 407--436.

\bibitem[\protect\citeauthoryear{McKinnon}{McKinnon}{2006}]{mckinnon2006review}
{\sc McKinnon, A.} 2006.
\newblock A review of {E}uropean truck tolling schemes and assessment of their
  possible impact on logistics systems.
\newblock {\em International Journal of Logistics\/}~{\em 9,\/}~3, 191--205.

\bibitem[\protect\citeauthoryear{McKnight and Bailey}{McKnight and
  Bailey}{1997}]{McKnight}
{\sc McKnight, L.~W.} {\sc and} {\sc Bailey, J.~P.} 1997.
\newblock An introduction to {I}nternet economics.
\newblock In {\em Internet Economics}, {L.~W. McKnight} {and} {J.~P. Bailey},
  Eds. {The MIT Press}, Cambridge, MA, 91--118.

\bibitem[\protect\citeauthoryear{Mohsenian-Rad and Leon-Garcia}{Mohsenian-Rad
  and Leon-Garcia}{2010}]{mohsenian2010optimalres}
{\sc Mohsenian-Rad, A.~H.} {\sc and} {\sc Leon-Garcia, A.} 2010.
\newblock Optimal residential load control with price prediction in real-time
  electricity pricing environments.
\newblock {\em IEEE Transactions on Smart Grid\/}~{\em 1,\/}~2, 120--133.

\bibitem[\protect\citeauthoryear{Mohsenian-Rad, Wong, Jatskevich, and
  Schober}{Mohsenian-Rad et~al\mbox{.}}{2010}]{mohsenian2010optimal}
{\sc Mohsenian-Rad, A.~H.}, {\sc Wong, V. W.~S.}, {\sc Jatskevich, J.}, {\sc
  and} {\sc Schober, R.} 2010.
\newblock Optimal and autonomous incentive-based energy consumption scheduling
  algorithm for smart grid.
\newblock In {\em Proceedings of the IEEE Conference on Innovative Smart Grid
  Technologies}. IEEE, 1--6.

\bibitem[\protect\citeauthoryear{Morgan}{Morgan}{2011}]{AsiaPricing}
{\sc Morgan, M.} 2011.
\newblock Pricing schemes key in {LTE} future.
\newblock Telecomasia.net.
\newblock September 12, http://www.
  telecomasia.net/content/pricing-schemes-key-lte-future.

\bibitem[\protect\citeauthoryear{Muller}{Muller}{2012}]{SA-sat}
{\sc Muller, R.} 2012.
\newblock {Satellite broadband pricing comparison}.
\newblock January 25, http://mybroadband.co.za/news/ broadband/.

\bibitem[\protect\citeauthoryear{Murphy and Murphy}{Murphy and
  Murphy}{1994}]{Murphy-Murphy}
{\sc Murphy, J.} {\sc and} {\sc Murphy, L.} 1994.
\newblock Bandwidth allocation by pricing in {ATM} networks.
\newblock {\em {IFIP} Transactions C: Communications Systems\/}~{\em C-24},
  333--351.

\bibitem[\protect\citeauthoryear{Murphy, Murphy, and Posner}{Murphy
  et~al\mbox{.}}{1994}]{Murphy-Posner}
{\sc Murphy, J.}, {\sc Murphy, L.}, {\sc and} {\sc Posner, E.~C.} 1994.
\newblock Distributed pricing for embedded {ATM} networks.
\newblock In {\em Proceedings of the International Teletraffic Conference}.
  Elsevier, 1053--1063.

\bibitem[\protect\citeauthoryear{Musacchio and Walrand}{Musacchio and
  Walrand}{2006}]{Musacchio}
{\sc Musacchio, J.} {\sc and} {\sc Walrand, J.} 2006.
\newblock {WiFi} access point pricing as a dynamic game.
\newblock {\em {IEEE/ACM} Transactions on Networking\/}~{\em 14}, 289--301.

\bibitem[\protect\citeauthoryear{Nash}{Nash}{1950}]{nash1950bargaining}
{\sc Nash, J.~F.} 1950.
\newblock {The bargaining problem}.
\newblock {\em Econometrica\/}~{\em 18,\/}~2, 155--162.

\bibitem[\protect\citeauthoryear{Neely, Modiano, and Li}{Neely
  et~al\mbox{.}}{2005}]{neely2005fairness}
{\sc Neely, M.}, {\sc Modiano, E.}, {\sc and} {\sc Li, C.} 2005.
\newblock Fairness and optimal stochastic control for heterogeneous networks.
\newblock In {\em Proceedings of IEEE INFOCOM}. Vol.~3. IEEE, 1723--1734.

\bibitem[\protect\citeauthoryear{Newman}{Newman}{2011}]{Newman}
{\sc Newman, J.} 2011.
\newblock Netflix has bandwidth cap sufferers covered.
\newblock PC World.
\newblock June 23,
  http://www.pcworld.com/article/230982/Netflix\_Has\_Bandwidth\_Cap\_Sufferers\_Covered.html.

\bibitem[\protect\citeauthoryear{Nicholson and Snyder}{Nicholson and
  Snyder}{2008}]{microeconomics}
{\sc Nicholson, W.} {\sc and} {\sc Snyder, C.} 2008.
\newblock {\em Microeconomic Theory\/} Tenth Ed.
\newblock {SouthWestern CENGAGE} Learning, Mason, OH.

\bibitem[\protect\citeauthoryear{Niyato and Hossain}{Niyato and
  Hossain}{2008}]{niyato2008competitive}
{\sc Niyato, D.} {\sc and} {\sc Hossain, E.} 2008.
\newblock Competitive pricing in heterogeneous wireless access networks: Issues
  and approaches.
\newblock {\em IEEE Network\/}~{\em 22,\/}~6, 4--11.

\bibitem[\protect\citeauthoryear{Odlyzko}{Odlyzko}{1999}]{Odlyzko}
{\sc Odlyzko, A.} 1999.
\newblock Paris metro pricing for the {I}nternet.
\newblock In {\em Proceedings of the 1st ACM Conference on Electronic
  Commerce}. ACM, 140--147.

\bibitem[\protect\citeauthoryear{Odlyzko}{Odlyzko}{2001}]{andrew}
{\sc Odlyzko, A.} 2001.
\newblock Internet pricing and the history of communications.
\newblock {\em Computer Networks\/}~{\em 36,\/}~5, 493--517.

\bibitem[\protect\citeauthoryear{Odlyzko, Arnaud, Stallman, and
  Weinberg}{Odlyzko et~al\mbox{.}}{2012}]{Odlyzko-Public}
{\sc Odlyzko, A.}, {\sc Arnaud, B.~S.}, {\sc Stallman, E.}, {\sc and} {\sc
  Weinberg, M.} 2012.
\newblock Know your limits: {C}onsidering the role of data caps and usage based
  billing in {I}nternet access service.
\newblock Public Knowledge.

\bibitem[\protect\citeauthoryear{Owoseje}{Owoseje}{2011}]{Swapables}
{\sc Owoseje, T.} 2011.
\newblock Orange refreshes {P}anther tariff with {S}wapables benefits.
\newblock Mobile Magazine.
\newblock September 7, http://tinyurl.com/Swapables.

\bibitem[\protect\citeauthoryear{Parker}{Parker}{2010}]{O2}
{\sc Parker, A.} 2010.
\newblock O2 scraps unlimited data for smartphones.
\newblock Financial Times.

\bibitem[\protect\citeauthoryear{Parris and Ferrari}{Parris and
  Ferrari}{1992}]{ParrisF}
{\sc Parris, C.} {\sc and} {\sc Ferrari, D.} 1992.
\newblock A resource based pricing policy for real-time channels in a
  packet-switching network.
\newblock Tech. rep., Tenet Group, ICSI, UC Berkeley.
\newblock TR-92-018.

\bibitem[\protect\citeauthoryear{Parris, Keshav, and Ferrari}{Parris
  et~al\mbox{.}}{1992}]{PKF}
{\sc Parris, C.}, {\sc Keshav, S.}, {\sc and} {\sc Ferrari, D.} 1992.
\newblock A framework for the study of pricing in integrated networks.
\newblock Tech. rep., Tenet Group, ICSI, UC Berkeley.
\newblock TR-92-016.

\bibitem[\protect\citeauthoryear{Paschalidis and Tsitsikilis}{Paschalidis and
  Tsitsikilis}{1998}]{Paschalidis}
{\sc Paschalidis, I.~C.} {\sc and} {\sc Tsitsikilis, J.~N.} 1998.
\newblock Congestion-dependent pricing of network services.
\newblock {\em IEEE/ACM Transactions on Networking\/}~{\em 8}, 171--184.

\bibitem[\protect\citeauthoryear{Poole and Orski}{Poole and Orski}{2003}]{hov}
{\sc Poole, R.} {\sc and} {\sc Orski, C.~K.} 2003.
\newblock {HOT} networks: A new plan for congestion relief and better transit.
\newblock Tech. rep., Reason Foundation. February.

\bibitem[\protect\citeauthoryear{{Princeton EDGE Lab}}{{Princeton EDGE
  Lab}}{2012}]{datawiz}
{\sc {Princeton EDGE Lab}}. 2012.
\newblock {DataWiz website}.
\newblock http://scenic.princeton.edu/datawiz/.

\bibitem[\protect\citeauthoryear{Reichl}{Reichl}{2010}]{reichl2010charging}
{\sc Reichl, P.} 2010.
\newblock From charging for quality of service to charging for quality of
  experience.
\newblock {\em Annals of Telecommunications\/}~{\em 65,\/}~3, 189--199.

\bibitem[\protect\citeauthoryear{Ren, Park, and van~der Schaar}{Ren
  et~al\mbox{.}}{2013}]{ren2012entry}
{\sc Ren, S.}, {\sc Park, J.}, {\sc and} {\sc van~der Schaar, M.} 2013.
\newblock Entry and spectrum sharing scheme selection in femtocell
  communications markets.
\newblock {\em IEEE/ACM Transactions on Networking\/}~{\em 21,\/}~1, 218--232.

\bibitem[\protect\citeauthoryear{Richards, Gilliam, and Larkinson}{Richards
  et~al\mbox{.}}{1996}]{richards1996london}
{\sc Richards, M.}, {\sc Gilliam, C.}, {\sc and} {\sc Larkinson, J.} 1996.
\newblock The {L}ondon congestion charging research programme. 1. {T}he
  progamme in overview.
\newblock {\em Traffic Engineering and Control\/}~{\em 37,\/}~2, 66--71.

\bibitem[\protect\citeauthoryear{Ricker}{Ricker}{2010}]{nycwifi}
{\sc Ricker, T.} 2010.
\newblock {AT\&T making tourists even more annoying with free Times Square
  WiFi}.
\newblock Engadget.
\newblock http://tinyurl.com/29quo58.

\bibitem[\protect\citeauthoryear{Rigney, Willens, Rubens, and Simpson}{Rigney
  et~al\mbox{.}}{2000}]{rfc2865}
{\sc Rigney, C.}, {\sc Willens, S.}, {\sc Rubens, A.}, {\sc and} {\sc Simpson,
  W.} 2000.
\newblock {Remote Authentication Dial In User Service (RADIUS)}.
\newblock RFC 2865 (Draft Standard).
\newblock Updated by RFCs 2868, 3575, 5080.

\bibitem[\protect\citeauthoryear{Roozbehani, Dahleh, and Mitter}{Roozbehani
  et~al\mbox{.}}{2010}]{roozbehani2010dynamic}
{\sc Roozbehani, M.}, {\sc Dahleh, M.}, {\sc and} {\sc Mitter, S.} 2010.
\newblock Dynamic pricing and stabilization of supply and demand in modern
  electric power grids.
\newblock In {\em Proceedings of the First IEEE International Conference on
  Smart Grid Communications}. IEEE, 543--548.

\bibitem[\protect\citeauthoryear{{Rural Broadband}}{{Rural
  Broadband}}{2012}]{avanti}
{\sc {Rural Broadband}}. 2012.
\newblock {Avanti Hylas 1 satellite broadband services}.
\newblock http://www.ruralbroadband.co.uk/ avanti-hylas-1-8mbps.html.

\bibitem[\protect\citeauthoryear{Sachson}{Sachson}{2011}]{tollfree}
{\sc Sachson, T.} 2011.
\newblock Subsidized data delivery via toll-free apps.
\newblock eComm 2011.

\bibitem[\protect\citeauthoryear{Samadi, Mohsenian-Rad, Schober, Wong, and
  Jatskevich}{Samadi et~al\mbox{.}}{2010}]{samadi2010optimal}
{\sc Samadi, P.}, {\sc Mohsenian-Rad, A.}, {\sc Schober, R.}, {\sc Wong, V.
  W.~S.}, {\sc and} {\sc Jatskevich, J.} 2010.
\newblock Optimal real-time pricing algorithm based on utility maximization for
  smart grid.
\newblock In {\em Proceedings of the First IEEE International Conference on
  Smart Grid Communications}. IEEE, 415--420.

\bibitem[\protect\citeauthoryear{Saraydar, Mandayam, and Goodman}{Saraydar
  et~al\mbox{.}}{2002}]{saraydar2002efficient}
{\sc Saraydar, C.}, {\sc Mandayam, N.}, {\sc and} {\sc Goodman, D.} 2002.
\newblock Efficient power control via pricing in wireless data networks.
\newblock {\em IEEE Transactions on Communications\/}~{\em 50,\/}~2, 291--303.

\bibitem[\protect\citeauthoryear{Sarkar}{Sarkar}{1997}]{Sarkar}
{\sc Sarkar, M.} 1997.
\newblock {I}nternet pricing: {A} regulatory imperative.
\newblock In {\em Internet Economics}, {L.~W. McKnight} {and} {J.~P. Bailey},
  Eds. {The MIT Press}, Cambridge, MA.

\bibitem[\protect\citeauthoryear{Schatz and Ante}{Schatz and Ante}{2010}]{FCC}
{\sc Schatz, A.} {\sc and} {\sc Ante, S.~E.} 2010.
\newblock {FCC} chief backs usage-based broadband pricing.
\newblock Wall Street Journal.
\newblock December 2.

\bibitem[\protect\citeauthoryear{Segall}{Segall}{2011}]{Verizon-plan}
{\sc Segall, L.} 2011.
\newblock Verizon ends unlimited data plan.
\newblock CNN Money.
\newblock July 6.

\bibitem[\protect\citeauthoryear{Semret, Liao, Campbell, and Lazar}{Semret
  et~al\mbox{.}}{2000}]{Semret}
{\sc Semret, N.}, {\sc Liao, R. R.-F.}, {\sc Campbell, A.~T.}, {\sc and} {\sc
  Lazar, A.} 2000.
\newblock Pricing, provisioning and peering: {D}ynamic markets for
  differentiated {I}nternet services and implications for network
  interconnections.
\newblock {\em {IEEE} Journal on Selected Areas in Communications\/}~{\em 18},
  2499--2513.

\bibitem[\protect\citeauthoryear{Sen}{Sen}{2011}]{Uninor}
{\sc Sen, S.} 2011.
\newblock {Bare-knuckled wireless}.
\newblock {Business Today}.
\newblock March 10.

\bibitem[\protect\citeauthoryear{Sen, Joe-Wong, and Ha}{Sen
  et~al\mbox{.}}{2012a}]{WITS}
{\sc Sen, S.}, {\sc Joe-Wong, C.}, {\sc and} {\sc Ha, S.} 2012a.
\newblock The economics of shared data plans.
\newblock In {\em Proceedings of 22nd Annual Workshop on Information
  Technologies and Systems (WITS)}.

\bibitem[\protect\citeauthoryear{Sen, Joe-Wong, Ha, Bawa, and Chiang}{Sen
  et~al\mbox{.}}{2013}]{sigchi}
{\sc Sen, S.}, {\sc Joe-Wong, C.}, {\sc Ha, S.}, {\sc Bawa, J.}, {\sc and} {\sc
  Chiang, M.} 2013.
\newblock When the price is right: Enabling time-dependent pricing of mobile
  data.
\newblock In {\em Proceedings of ACM SIGCHI}. ACM.

\bibitem[\protect\citeauthoryear{Sen, Joe-Wong, Ha, and Chiang}{Sen
  et~al\mbox{.}}{2012b}]{sen2012incentivizing}
{\sc Sen, S.}, {\sc Joe-Wong, C.}, {\sc Ha, S.}, {\sc and} {\sc Chiang, M.}
  2012b.
\newblock Incentivizing time-shifting of data: {A} survey of time-dependent
  pricing for {I}nternet access.
\newblock {\em IEEE Communications Magazine\/}~{\em 50,\/}~11, 91--99.

\bibitem[\protect\citeauthoryear{Sephton}{Sephton}{2011a}]{ContentPricing}
{\sc Sephton, J.} 2011a.
\newblock The changing pricing model for mobile content.
\newblock Strategy Analytics.
\newblock November 24, http://tinyurl.com/ContentPricing.

\bibitem[\protect\citeauthoryear{Sephton}{Sephton}{2011b}]{Spain}
{\sc Sephton, J.} 2011b.
\newblock Spain - {N}ew {O}range {D}elfin price plan.
\newblock Strategy Analytics.
\newblock July 13, http://tinyurl.com/ OrangeSpain.

\bibitem[\protect\citeauthoryear{Sephton}{Sephton}{2011c}]{VodafoneTest}
{\sc Sephton, J.} 2011c.
\newblock Trialling mobile data - a new trend for mobile plans?
\newblock Strategy Analytics.
\newblock October 28, http://tinyurl.com/mobiletrials.

\bibitem[\protect\citeauthoryear{Shakkotai, Srikant, Ozdaglar, and
  Acemoglu}{Shakkotai et~al\mbox{.}}{2008}]{shakkotai}
{\sc Shakkotai, S.}, {\sc Srikant, R.}, {\sc Ozdaglar, A.}, {\sc and} {\sc
  Acemoglu, D.} 2008.
\newblock The price of simplicity.
\newblock {\em {IEEE Journal on Selected Areas in Communication}\/}~{\em
  26,\/}~7, 1269--1276.

\bibitem[\protect\citeauthoryear{Shen and Basar}{Shen and Basar}{2007}]{Basar}
{\sc Shen, H.} {\sc and} {\sc Basar, T.} 2007.
\newblock Optimal nonlinear pricing for a monopolistic network service provider
  with complete and incomplete information.
\newblock {\em {IEEE} Journal on Selected Areas in Communication\/}~{\em 25},
  1216--1223.

\bibitem[\protect\citeauthoryear{Sheng, Joe-Wong, Ha, Wong, and Sen}{Sheng
  et~al\mbox{.}}{2013}]{workshop}
{\sc Sheng, M.-J.}, {\sc Joe-Wong, C.}, {\sc Ha, S.}, {\sc Wong, F. M.~F.},
  {\sc and} {\sc Sen, S.} 2013.
\newblock Smart data pricing: {L}essons from trial planning.
\newblock In {\em Proceedings of the Second Smart Data Pricing Workshop}. IEEE.

\bibitem[\protect\citeauthoryear{Shenker, Clark, Estrin, and Herzog}{Shenker
  et~al\mbox{.}}{1996}]{Shenker96}
{\sc Shenker, S.}, {\sc Clark, D.}, {\sc Estrin, D.}, {\sc and} {\sc Herzog,
  S.} 1996.
\newblock Pricing in computer networks: Reshaping the research agenda.
\newblock {\em Telecommunications Policy\/}~{\em 20,\/}~3, 183--201.

\bibitem[\protect\citeauthoryear{Shetty, Parekh, and Walrand}{Shetty
  et~al\mbox{.}}{2009}]{shetty2009economics}
{\sc Shetty, N.}, {\sc Parekh, S.}, {\sc and} {\sc Walrand, J.} 2009.
\newblock Economics of femtocells.
\newblock In {\em Proceedings of IEEE GLOBECOM}. IEEE, 1--6.

\bibitem[\protect\citeauthoryear{Songhurst}{Songhurst}{1999}]{Songhurst}
{\sc Songhurst, D.} 1999.
\newblock {\em Charging communication networks: From theory to practice}.
\newblock Elsevier.

\bibitem[\protect\citeauthoryear{Sun and Modiano}{Sun and
  Modiano}{2006}]{sun2006channel}
{\sc Sun, J.} {\sc and} {\sc Modiano, E.} 2006.
\newblock Channel allocation using pricing in satellite networks.
\newblock In {\em Proceedings of the 40th Annual Conference on Information
  Sciences and Systems}. IEEE, 182--187.

\bibitem[\protect\citeauthoryear{Svitak}{Svitak}{2012}]{svitak}
{\sc Svitak, A.} 2012.
\newblock {Demand for bandwidth drives commercial {S}atcom}.
\newblock Aviation Week.
\newblock January 30.

\bibitem[\protect\citeauthoryear{Taylor}{Taylor}{2000}]{rfc2886}
{\sc Taylor, T.} 2000.
\newblock {Megaco Errata}.
\newblock RFC 2886 (Historic).
\newblock Obsoleted by RFC 3015.

\bibitem[\protect\citeauthoryear{Teitell}{Teitell}{2012}]{BostonGlobe}
{\sc Teitell, B.} 2012.
\newblock {Cellphone overcharges putting a strain on many families}.
\newblock Boston Globe.
\newblock September 22.

\bibitem[\protect\citeauthoryear{Telecompaper}{Telecompaper}{2010}]{Mobinil}
{\sc Telecompaper}. 2010.
\newblock Mobinil offers time-based mobile data plans.
\newblock July 19, http://www.telecompaper.com/
  news/mobinil-offers-timebased-mobile-data-plans.

\bibitem[\protect\citeauthoryear{{The Economist}}{{The
  Economist}}{2009}]{Economist}
{\sc {The Economist}}. 2009.
\newblock {The mother of invention: Network operators in the poor world are
  cutting costs and increasing access in innovative ways}.
\newblock {Special Report, September 24}.

\bibitem[\protect\citeauthoryear{{U.S. EIA}}{{U.S. Energy Information Administration}}{2012}]{EIA}
{\sc {U.S. Energy Information Administration}}. 2012.
\newblock Wholesale market data.
\newblock http://www.eia.gov/electricity/wholesale/.

\bibitem[\protect\citeauthoryear{{U.S. Office of Highway Policy
  Information}}{{U.S. Office of Highway Policy Information}}{2011}]{ushighways}
{\sc {U.S. Office of Highway Policy Information}}. 2011.
\newblock Toll facilities in the {U}nited {S}tates.
\newblock Publication No: FHWA-PL-11-032.

\bibitem[\protect\citeauthoryear{Varaiya}{Varaiya}{1999}]{varaiya1999pricing}
{\sc Varaiya, P.} 1999.
\newblock Pricing and provisioning of quality-differentiated services.
\newblock In {\em Proceedings of the Information Theory and Networking
  Workshop}. IEEE, 38.

\bibitem[\protect\citeauthoryear{Vardy}{Vardy}{2011}]{Telus}
{\sc Vardy, M.} 2011.
\newblock Telus tuneage: Now powered by {R}dio.
\newblock The Next Web.
\newblock August 3, http://thenextweb.com/ca/
  2011/08/03/telus-tuneage-now-powered-by-rdio/.

\bibitem[\protect\citeauthoryear{Vytelingum, Ramchurn, Voice, Rogers, and
  Jennings}{Vytelingum et~al\mbox{.}}{2010}]{vytelingum2010trading}
{\sc Vytelingum, P.}, {\sc Ramchurn, S.~D.}, {\sc Voice, T.~D.}, {\sc Rogers,
  A.}, {\sc and} {\sc Jennings, N.~R.} 2010.
\newblock Trading agents for the smart electricity grid.
\newblock In {\em Proceedings of the 9th International Conference on Autonomous
  Agents and Multiagent Systems: Volume 1}. International Foundation for
  Autonomous Agents and Multiagent Systems, 897--904.

\bibitem[\protect\citeauthoryear{Wagner, {Van den Berg}, Giacopelli, Ghetie,
  Burns, Tauil, Sen, Wang, Chiang, Lan, Laddaga, Robertson, and
  Manghwani}{Wagner et~al\mbox{.}}{2012}]{Wagner}
{\sc Wagner, S.}, {\sc {Van den Berg}, E.}, {\sc Giacopelli, J.}, {\sc Ghetie,
  A.}, {\sc Burns, J.}, {\sc Tauil, M.}, {\sc Sen, S.}, {\sc Wang, M.}, {\sc
  Chiang, M.}, {\sc Lan, T.}, {\sc Laddaga, R.}, {\sc Robertson, P.}, {\sc and}
  {\sc Manghwani, P.} 2012.
\newblock Autonomous, collaborative control for resilient cyber defense
  ({ACCORD}).
\newblock In {\em Proceedings of the Workshop on Adaptive Host and Network
  Security (AHANS)}. IEEE.

\bibitem[\protect\citeauthoryear{Wahlmueller, Zwickl, and Reichl}{Wahlmueller
  et~al\mbox{.}}{2012}]{wahlmueller2012pricing}
{\sc Wahlmueller, S.}, {\sc Zwickl, P.}, {\sc and} {\sc Reichl, P.} 2012.
\newblock Pricing and regulating quality of experience.
\newblock In {\em Proceedings of the 8th EURO-NF Conference on Next Generation
  Internet}. IEEE, 57--64.

\bibitem[\protect\citeauthoryear{Walrand}{Walrand}{2008}]{walrand}
{\sc Walrand, J.} 2008.
\newblock Economic models of communication networks.
\newblock In {\em Performance Modeling and Engineering}, {Z.~Liu} {and} {C.~H.
  Xia}, Eds. Springer Publishing Company, New York, Chapter~3, 57--90.

\bibitem[\protect\citeauthoryear{Wells and Haas}{Wells and
  Haas}{2004}]{wells2004electricity}
{\sc Wells, J.} {\sc and} {\sc Haas, D.} 2004.
\newblock {\em Electricity markets: Consumers could benefit from demand
  programs, but challenges remain}.
\newblock DIANE Publishing, Darby, PA.

\bibitem[\protect\citeauthoryear{Wolak}{Wolak}{2006}]{wolak2006residential}
{\sc Wolak, F.} 2006.
\newblock Residential customer response to real-time pricing: {T}he {A}naheim
  critical-peak pricing experiment.
\newblock Tech. rep., Stanford University.
\newblock http://www.stanford.edu/wolak.

\bibitem[\protect\citeauthoryear{Ya\"{i}che, Mazumdar, and
  Rosenberg}{Ya\"{i}che et~al\mbox{.}}{2000}]{Yaiche}
{\sc Ya\"{i}che, H.}, {\sc Mazumdar, R.~R.}, {\sc and} {\sc Rosenberg, C.}
  2000.
\newblock A game theoretic framework for bandwidth allocation and pricing in
  broadband networks.
\newblock {\em {IEEE/ACM} Transactions on Networking\/}~{\em 8}, 667--678.

\bibitem[\protect\citeauthoryear{Yoo}{Yoo}{2009}]{Yoo}
{\sc Yoo, C.~S.} 2009.
\newblock Network neutrality, consumers, and innovation.
\newblock {\em University of Chicago Legal Forum\/}~{\em 25}, 179.
\newblock U of Penn Law School, Public Law Research Paper No. 08-40.

\end{thebibliography}


  
\end{document}